\documentclass[twocolumn,trackchanges]{aastex631} %for Arxiv format

\usepackage{amsmath,amssymb}
\usepackage{array}
\usepackage{enumitem} 
\usepackage{rotating} 
\usepackage{booktabs} 
\usepackage{comment} 

\newcommand{\met}{$\mathrm{CH_4}$ }

\begin{document}

\title{Measurement of Methane Line Broadening in Hot Hydrogen/Helium Atmospheres at $\lambda$~=~1.60--1.63~{\textmu}m for Substellar Object Spectroscopy}

\author[0000-0001-5414-4171]{Ko Hosokawa}
\affiliation{Astronomical Science Program, The Graduate University for Advanced Studies, SOKENDAI, \\2-21-1 Osawa, Mitaka, Tokyo 181-8588, Japan}
\email{ko.hosokawa.astro@gmail.com}

\author[0000-0001-6181-3142]{Takayuki Kotani}
\affiliation{Astrobiology Center, 2-21-1 Osawa, Mitaka, Tokyo 181-8588, Japan}
\affiliation{National Astronomical Observatory of Japan, 2-21-1 Osawa, Mitaka, Tokyo 181-8588, Japan}
\affiliation{Astronomical Science Program, The Graduate University for Advanced Studies, SOKENDAI, \\2-21-1 Osawa, Mitaka, Tokyo 181-8588, Japan}

\author[0000-0003-3309-9134]{Hajime Kawahara}

\affiliation{Department of Space Astronomy and Astrophysics, ISAS/JAXA, 3-1-1, Yoshinodai, Sagamihara, Kanagawa, 252-5210 Japan}
\affiliation{Department of Astronomy, Graduate School of Science, The University of Tokyo, 7-3-1 Hongo, Bunkyo-ku, Tokyo 113-0033, Japan}

\author[0000-0003-3800-7518]{Yui Kawashima}
\affiliation{Department of Astronomy, Graduate School of Science, Kyoto University, Kitashirakawa Oiwake-cho, Sakyo-ku, Kyoto 606-8502, Japan}
\affiliation{Frontier Research Institute for Interdisciplinary Sciences and Department of Geophysics, Tohoku University, 
\\Aramaki aza Aoba 6-3, Aoba-ku, Sendai, 980-8578, Japan}
\affiliation{Department of Space Astronomy and Astrophysics, ISAS/JAXA, 3-1-1, Yoshinodai, Sagamihara, Kanagawa, 252-5210 Japan}
\affiliation{Cluster for Pioneering Research, RIKEN, 2-1 Hirosawa, Wako, Saitama 351-0198, Japan}

\author[0000-0003-1298-9699]{Kento Masuda}
\affiliation{Department of Earth and Space Science, Osaka University, 1-1 Machikaneyama-cho, Toyonaka-shi, Osaka 560-0043, Japan}

\author[0000-0003-3881-3202]{Aoi Takahashi}
\affiliation{Department of Space Astronomy and Astrophysics, ISAS/JAXA, 3-1-1, Yoshinodai, Sagamihara, Kanagawa, 252-5210 Japan}

\author[0000-0001-5451-9367]{Kazuo Yoshioka}
\affiliation{Department of Complexity Science and Engineering, The University of Tokyo, 5-1-5 Kashiwanoha, Kashiwa, Chiba 277-8561, Japan}


\begin{abstract}
Recent high-dispersion spectroscopy from ground-based telescopes and high-precision spectroscopy from space observatories have enabled atmospheric observations of substellar objects, such as brown dwarfs and hot gaseous exoplanets, with sufficient precision to make ambient gas differences in molecular line broadening a significant factor. In this paper, we experimentally measured the pressure broadening of methane in a high-temperature  hydrogen-helium background atmosphere in the H band, which had not been previously measured. The experiment used glass cells, inserted in a tube furnace, filled with methane in a hydrogen-helium background atmosphere or pure methane gas. Spectra were obtained at four temperatures ranging from room temperature to 1000~K, in the wavelength range 1.60--1.63~{\textmu}m, using a tunable laser, yielding eight high-resolution spectra in total. A full Bayesian analysis was performed on the obtained spectra, using the differentiable spectral model {\sf ExoJAX} and the Hamiltonian Monte Carlo for inferring a large number of parameters, allowing us to infer the $\mathrm{H}_2$/He pressure broadening for 22 transitions mainly in the R-branch of the 2$\nu_3$ band. 
As a result, we found a temperature exponent of approximately 0.27 
and a reference width at 296~K of around 0.040 for \( J_\mathrm{lower} \) = 13--20. This temperature dependency is much milder than that provided by the molecular database ExoMol, yielding a line width approximately 5--45\% smaller than ExoMol at 296~K, but similar at 1000~K. Our results suggest the need for further accumulation of experimental data for spectral analysis of substellar objects with hydrogen-helium atmospheres.
\end{abstract}

\section{Introduction} \label{sec:intro}
Thanks to recent advances in observational technology, thousands of substellar objects, including planets and brown dwarfs, have been discovered.
These substellar objects exhibit a wide range of masses, radii, and orbital periods, including parameter spaces that have no counterparts in our solar system.
To understand the origin of this diversity, spectroscopic observations of their atmospheres are crucial since atmospheric properties, such as elemental abundances and temperature structures, reflect their formation and evolutionary processes \citep[][]{2011ApJ...743L..16O, 2002A&A...385..156G}.\par
To infer the atmospheric properties of the observed targets, it is essential to compare observed spectra with theoretical models.
The accuracy of these models depends heavily on molecular line list databases, which provide critical information for each absorption line, such as central wavelength, line strength, and broadening parameters.
A major challenge in the characterization of exoplanet atmospheres stems from the incompleteness of line list databases that are suitable for their significantly different environment from the Earth.
Due to ease of observation, the current primary targets are gas giants with short or long orbital periods, whose atmospheres are dominated by $\mathrm{H_2}$ and $\mathrm{He}$, and have temperatures of $\gtrsim 500$~K.
While most molecular line databases, such as HITRAN \citep{1987ApOpt..26.4058R} and GEISA \citep{1986AnGeo...4..185H}, have been developed for the environment of the Earth, namely the $\mathrm{N_2}$-dominated atmosphere at $\sim 300$~K, the difference in temperature and background atmosphere greatly matters since the energy states and pressure-broadening depend on temperature and colliding species in the background atmosphere.\par
In response to the increasing demand, molecular line list databases tailored for high-temperature ($\gtrsim 1000$~K) and $\mathrm{H_2}$-rich environments have started to be developed through projects such as HITEMP \citep{2010JQSRT.111.2139R} and ExoMol \citep{2016JMoSp.327...73T}.
\cite{niraula_impending_2022} recently highlighted the substantial impact of line list databases and our limited knowledge about line profiles on inferring atmospheric properties from the observed spectra of exoplanets. 
In their study, 
they examined nine opacity models, each utilizing different pressure-broadening parameters, far-wing line profile treatments, line list databases (HITRAN/ExoMol/HITEMP), or adjustments to line list parameters within measurement errors.
Their results showed that even at a spectral resolution of $\sim 1000$, which is insufficient to resolve individual line profiles, uncertainties in line list parameters and limitations in our understanding of line profiles restrict the accuracy of retrieving the atmospheric properties to $\sim 0.5$--$1.0$~dex, corresponding to factors of 3--10 times.\par
Among the various parameters in the line list database, pressure-broadening parameters are particularly important.
Pressure broadening refers to the broadening of spectral lines due to modifications in energy states through the collisional interaction with background molecules.
This effect becomes more prominent at high pressures, meaning that it generally dominates the broadening at the line wings, and the degree of broadening depends on the type of colliding molecular species.
An accurate knowledge of pressure broadening is essential to understand the atmospheres of observed targets.
For instance, uncertainties in broadening parameters limit the precision of gravity inference through the high-resolution line profile measurement, which is particularly important for planets at wide separation whose mass measurement via radial velocity is challenging.
In fact, \citet{kawahara_autodifferentiable_2022} demonstrated that the differences in the broadening parameters between the ExoMol and HITEMP databases for $\mathrm{CO}$ significantly impact mass inference from the high resolution spectrum of a brown dwarf.
Furthermore, pressure broadening plays a crucial role in determining the surface temperatures of Earth-like planets \citep{2009NatGe...2..891G}.
Or, in principle, with complete knowledge of pressure broadening, it would be possible to infer the composition of the background molecules from line profile measurements, even when the background molecules are radiatively inactive, such as $\mathrm{H_2}$ and $\mathrm{N_2}$.\par
As noted above, most line list databases are developed for terrestrial atmosphere, meaning that their pressure-broadening parameters are designed for an $\mathrm{N_2}$-dominated atmosphere at $\sim 300$~K.
However, the ExoMol database is tailored for an $\mathrm{H_2}$/$\mathrm{He}$-dominated atmosphere and includes the corresponding pressure-broadening parameters.
These parameters are provided for each quantum number assignment and consist of experimentally measured values compiled from the literature, covering full or partial quantum number assignments.
In the case of the HITRAN / HITEMP databases, the $\mathrm{H_2}$-, $\mathrm{He}$- and $\mathrm{CO_2}$-broadening parameters, along with their temperature dependencies, are included for some molecules, based on semiempirical models derived from both experimental measurements and theoretical predictions
\citep{2015ApJS..216...15L, 2016JQSRT.168..193W, 2022ApJS..262...40T}.

\begin{deluxetable*}{lllll}
\tablecaption{Previous measurements of $\mathrm{H_2}$ and $\mathrm{He}$ pressure broadening parameters for $\mathrm{CH_4}$, listed in order of publication year}
%The order is based on the publicated year. 
\label{tab:previous_exp}
%\tablewidth{0pt}
\tablehead{
\colhead{Reference} & \colhead{Broadener} & \colhead{Temperature [K]} & \colhead{Wavelength [{\textmu}m]} & \colhead{$J_\mathrm{lower}$}}

%\decimalcolnumbers
\startdata
$\ast$\cite{varanasi_collision-broadened_1971} & $\mathrm{H_2}$, He & 295 & 3.13--3.15, 3.27--3.29 & 0--6, 14--16 \\
$\ast$\cite{varanasi_experimental_1972} & $\mathrm{H_2}$, He & 295 & 7.56--8.28&0--2\\
\cite{darnton_temperature_1973}&$\mathrm{H_2}$, He &100, 298&1.67&0--2\\
\cite{keffer_pressure_1986} & $\mathrm{H_2}$, He & 77--295 & 0.61--0.68 & - \\
$\ast$\cite{fox_measurements_1988} & $\mathrm{H_2}$, He & 296 & $\sim$ 1.1 & 0--6\\
$\ast$\cite{varanasi_measurements_1989} & $\mathrm{H_2}$, He & 130--295 & 7.59--8.03 & 0, 1, 5, 9\\
$\ast$\cite{varanasi_temperature_1990} & $\mathrm{H_2}$, He & 161--295 & 7.40--7.50 &4, 7\\
$\ast$\cite{pine_self-_1992} & $\mathrm{H_2}$ & 296 & 3.31--3.32 &1--13\\
\cite{margolis_hydrogen_1993} & $\mathrm{H_2}$ & 297 & 2.33--2.42 &1--12\\
$\ast$\cite{grigoriev_estimation_2001} & He & 296 & 3.13--3.47 &0--13\\
\cite{pine_multispectrum_2003} & $\mathrm{H_2}$, He & 296 & 3.31--3.32 &1--13\\
$\ast$\cite{gabard_helium_2004} & He & 296 & 5.71--7.14 &2--16\\
\cite{gharib-nezhad_h2-induced_2019} & $\mathrm{H_2}$ & 300, 475, 655 & 3.13--3.57 &2--17\\
\cite{pine_speed-dependent_2019} & He & 296 & 3.31--3.32 &1--13\\
\cite{sung_h2-pressure_2020} & $\mathrm{H_2}$ & 80--370 & 2.16--2.22 &0--6\\
\cite{yousefi_line_2021} & $\mathrm{H_2}$ & 296--1098 & 3.12--3.63 &0--16\\
\cite{es-sebbar_line-strengths_2021} & $\mathrm{H_2}$, He & 297 & 3.37--3.47 &5--13\\
\enddata
\tablecomments{References with $\ast$ are used for the calculation of the broadening parameters in ExoMol by \cite{barton_exomol_2017}. For $J_\mathrm{lower}$, "-" is displayed if there is no clear explanation in the reference. }
\end{deluxetable*}
\vspace{-0.8cm}

Experimental measurements of the pressure-broadening parameter and its temperature dependency for an $\mathrm{H_2}$/$\mathrm{He}$-dominated atmosphere are generally limited.
Since pressure-broadening parameters can vary with vibrational and rotational quantum numbers, it is essential to measure these parameters for each vibrational band.
In this paper, we present experimental measurements of the pressure-broadening parameter and its temperature dependency in a high-temperature (up to 1000~K) hydrogen-helium background atmosphere.
We focus on methane because of its significant importance for substellar atmospheres and the current lack of experimental data.
\met is expected to be the dominant carbon-bearing species at $\lesssim 1000$~K in atmospheres with solar-like elemental abundances.
Indeed, it shapes the atmospheric spectra of brown dwarfs \citep[e.g.,][]{1995Sci...270.1478O} and directly imaged planets \citep[e.g.,][]{2013ApJ...778L...4J}.
In addition, its first robust detection in a transiting planet was recently reported based on the JWST/NIRCam observations \citep{2023Natur.623..709B}.
We compile previous representative measurements of the $\mathrm{H_2}$- and $\mathrm{He}$-broadening parameters of $\mathrm{CH_4}$ in Table~\ref{tab:previous_exp}.
Among these studies, \cite{gharib-nezhad_h2-induced_2019} and \cite{yousefi_line_2021} conducted measurements at high temperatures to simulate exoplanet atmospheres.
Both studies focused on the $\nu_3$ band at 3.3~{\textmu}m.
Based on fitting the measured data, \cite{gharib-nezhad_h2-induced_2019} provided a formula for the broadening parameters as a function of a quantum number, which is suitable for high-temperature $\mathrm{H_2}$-dominated atmospheres.
In this study, we selected the wavelength range of 1.60--1.63~{\textmu}m, which corresponds to a region where the R-branch of 2$\nu_3$ band in the fourth vibrational polyad ($P_4$) of $\mathrm{CH_4}$ dominates the spectra of relatively cool ($\lesssim 1000$~K) atmospheres with solar-like elemental abundances, such as those found in brown dwarfs and gas giant planets \citep[e.g.,][]{2022MNRAS.514.3160T, 2024arXiv241011561K}.
This range falls within the $H$-band, a typical band used for exoplanet atmospheric characterization by high-resolution spectrographs mounted on 10~m-class ground telescopes, such as VLT/CRIRES+ \citep{2004SPIE.5492.1218K, 2014SPIE.9147E..19F}, Gemini South/IGRINS \citep{2010SPIE.7735E..1MY, 2014SPIE.9147E..1DP, 2018SPIE10702E..0QM}, and Subaru/IRD \citep{2012SPIE.8446E..1TT, 2018SPIE10702E..11K}.
As shown in Table~\ref{tab:previous_exp}, there are no previous measurements of the $\mathrm{H_2}$-and $\mathrm{He}$-broadening parameter within the wavelength range of interest at high temperatures.

The remainder of this paper is organized as follows. In Section~\ref{sec:experiment}, we describe our experimental setup, procedures, and data reduction process.
We then present the obtained spectra and compare them with a previous study for verification.
Section~\ref{sec:modeling} details our analysis of the spectra using Bayesian modeling to derive a pressure-broadening parameter and its temperature dependency, followed by the presentation of the results in Section~\ref{sec:result}.
Finally, we conclude with a discussion and summary in Section~\ref{sec:summary}.

% Replace whole section of Experiment by Takayuki Kotani 2024/11/04
\section{Experiment}\label{sec:experiment}
\subsection{Experimental Setup}\label{sec:experiment_setup}
This section will describe the experimental setup used to measure the broadening effect of CH\(_4\)  absorption lines in the solar abundance H\(_2\) and He atmosphere, along with the method of data reduction. The experiment was executed at the JAXA Akiruno Experiment Laboratory in Tokyo, a facility designed with anti-explosion features for safety considerations.

\begin{figure*}[t]
\centering
\includegraphics[width=0.6\linewidth]{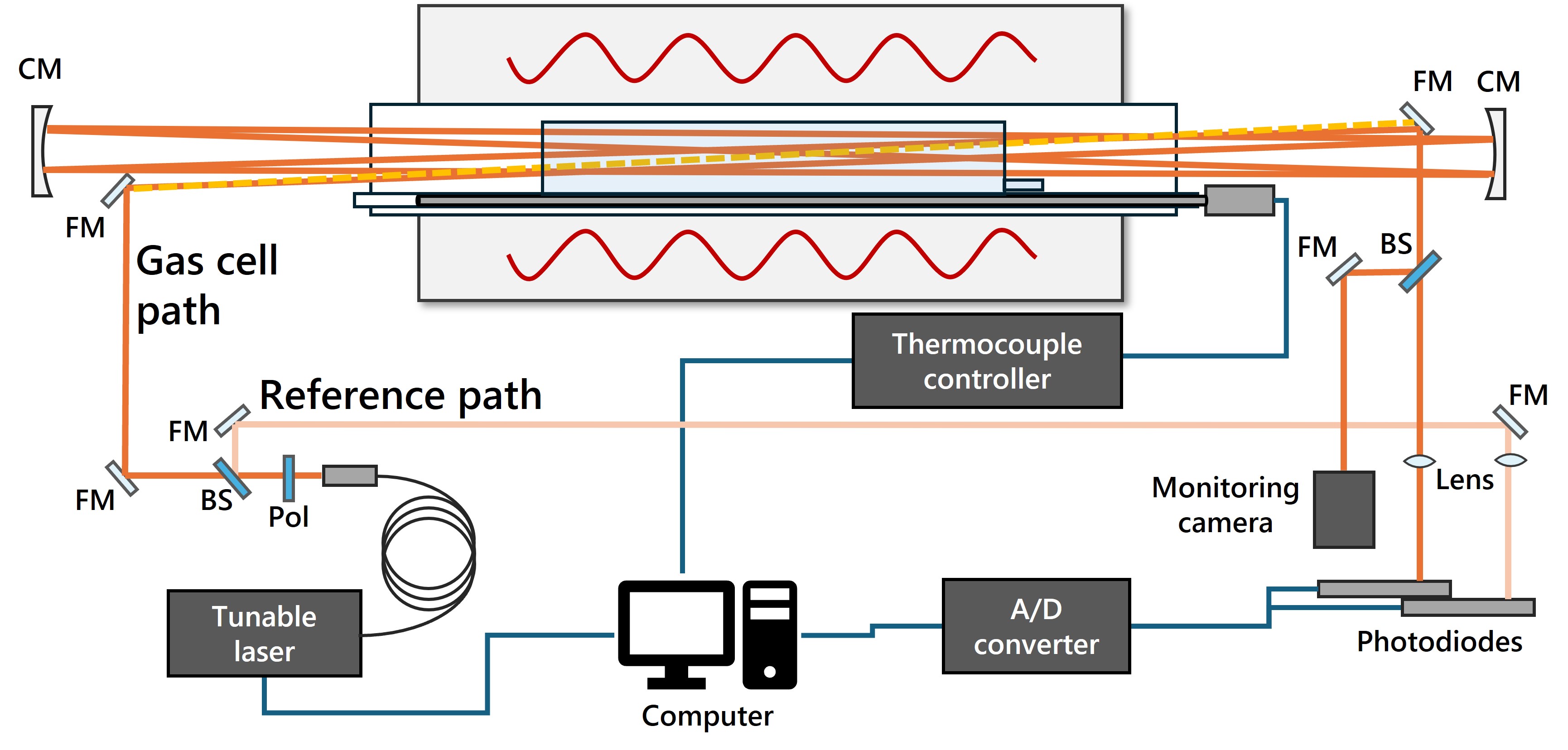}
\caption{Schematic of the experimental system. The dashed line indicates the single-path transmission case. Pol: Polarizer, FM: Flat mirror, CM: Concave mirror, BS: Beam splitter}
\label{fig:ExperimentScheme}
\end{figure*}

Figure~\ref{fig:ExperimentScheme} presents the schematic of the experimental setup. The light source is a tunable laser operating in the wavelength range of 1525 to 1630~nm, with a linewidth of less than 100 kHz, a maximum spectral resolution of 0.1 pm, and an absolute wavelength accuracy of $\pm$ 10~pm. The laser power was set at 1~mW over the entire spectral wavelength range, and no significant changes in measured spectra occurred at other power settings. The laser is placed in an air-conditioned environment and the light is delivered through 20 meters of polarization-maintaining single-mode fibers. The output laser light is collimated using an achromatic lens with a focal length of 15~mm, resulting in a beam waist diameter of 2.9~mm. Due to the elliptical and slightly unstable polarization state of the laser light, it was passed through a linear polarizer to select a specific polarization state, thereby minimizing polarization-dependent measurement instability. A beam splitter divides the incoming beam into two separate paths. \par
The transmitted light was directed to a gas cell for the measurement of the absorption of the gas sealed within a glass enclosure, while the reflected beam bypassed the gas cell and reached a photodiode for measurement of the laser intensity. The gas cell was housed inside a tube-shaped electric furnace, enabling temperature adjustments from room temperature up to 1000~K. The temperature within the gas cell was monitored at 8~points, located 60~mm apart beneath the cell, using K-type thermocouples. The temperature variation across all measurement points was maintained within $\pm$ 15~K of the target temperature. \par
For the measurement of gas absorption, we used a Herriott cell configuration \citep{herriott_off-axis_1964}, which consists of two concave spherical mirrors at the ends of the gas cell. This setup allows for a long optical path, enhancing the sensitivity of absorption measurements through multi-pass transmission while minimizing the increase in beam diameter due to diffraction. In the case of a 10\% CH\(_4\) gas cell, we implemented a 5-path configuration, resulting in an optical path length of 2485 mm to achieve an appropriate absorption depth. For the 100\% CH\(_4\) gas cell, we used a 1-path configuration to prevent saturation of the absorption depth. Both beams, one transmitted through the gas cell and the other for the reference path, were focused onto separate photodiode sensors using an $f$=100 mm achromatic lens, enabling simultaneous intensity monitoring of the laser light. We refer to the spectrum passing through the gas cell as the raw spectrum, and the spectrum obtained from intensity monitoring as the reference spectrum. A pinhole mask with a diameter of 1.3 mm was placed in front of the photodiodes to block stray light. Additionally, the photodiodes were slightly tilted with respect to the optical axis to prevent reflected light from the sensor surface from returning to the photodiode.

\subsection{Gas cell}\label{sec:gascell}
We used two types of gas cells for this experiment (Table~\ref{tab:Gascell_spec}). The first is a gas cell filled with a mixture of H\(_2\), He, and 10\% CH\(_4\) (hereafter referred to as the CH\(_4\) 10\% gas cell). The second gas cell contains 100\% CH\(_4\). 
%These cells are used to measure the pressure broadening effects caused by H\(_2\) and He, as well as the self-broadening of CH\(_4\).
The former gas cell is primarily used for measuring H\(_2\)/He broadening, but since it contains 10\% methane, self-broadening effects are also present. Therefore, we use the latter 100\% methane gas to simultaneously measure the effect of self-broadening.
The volume mixing ratio of H\(_2\) and He was configured to match the protosolar abundance \citep{Lodders2003ApJ}. Both gas cells are cylindrical in shape, with a diameter of 40~mm and a 500~mm long, and 2~mm thick circular windows made of synthetic silica, which are bonded by optical contact to both ends of the cylinder. The windows are inclined at an angle of 1 degree relative to the cylinder axis. In the case of the CH\(_4\) 100\% gas cell, we used plane-parallel glass plates for the windows, which resulted in significant optical interference and this interference appeared as a sinusoidal pattern in the spectrum. For the CH\(_4\) 10\% gas cell, the windows have 0.5-degree wedges designed to minimize optical interference, and we did not observe any significant interference in this gas cell.

\subsection{Data acquisition and reduction}
The absorption spectrum was obtained using the following procedures: The heater temperature of the electric furnace was increased at a speed of 7~K/min to avoid the break of the gas cell due to thermal stress. Once the temperature reached the target temperature, we waited at least 30 minutes before starting spectrum data acquisition to minimize the temperature gradient across the gas cell and ensure temperature stabilization. The laser wavelength was set from 1600 to 1630~nm at 0.25 pm steps every 0.2 seconds, and the analog voltage signals of the two photodiode interfaces were recorded simultaneously through an A/D converter and a PC for each wavelength setting of the tunable laser. 

Spectra were obtained for eight setups, the CH\(_4\) 100\% and CH\(_4\) 10\% gas cells at four different temperatures: room temperature (297~K), 500~K, 700~K, and 1000~K. The internal pressure of the gas cell varies with temperature, as there is no mechanism for pressure adjustment. Detailed temperature and pressure conditions for the measurements are shown in Table ~\ref{tab:expTPcondition}. Additionally, we collected the spectra without the gas cell to perform wavelength dependence corrections. We refer to the raw spectrum without the gas cell divided by its reference spectrum as the null spectrum.

\vspace{0.5cm}
\begin{table}[h]
    \centering
    \caption{Gas cell specifications}
    \begin{tabular}{ccc}
        \hline\hline
        Cell ID & Composition & Sealed pressure (Temperature) \\
        \hline
        1 & CH\(_4\) 100\% & 0.423 bar (296.8~K) \\
        \addlinespace\addlinespace
         & H\(_2\) 75.79\% &  \\
        2  & He 14.40\% & 0.423 bar (294.6~K)\\
          & CH\(_4\) 9.81\% & \\
        \hline
    \end{tabular}
    
    \label{tab:Gascell_spec}
\end{table}

\begin{table}[ht]
\centering
\caption{Setup ID, cell ID, pressure, and target temperature for the eight setups}
\begin{tabular}{cccc}
\hline\hline
Setup ID & Cell ID & {Pressure } & {Target temperature} \\ 
$k$& & (bar) & (K) \\ \hline
1 & 1 & 0.43 & room ($296.6^{+0.2}_{-0.1})^\ast$ \\ 
2 & 1 & 0.71 & 500 \\ 
3 & 1 & 1.00 & 700 \\ 
4 & 1 & 1.43 & 1000 \\ 
5 & 2 & 0.43 & room ($297.0^{+0.1}_{-0.2})^\ast$ \\ 
6 & 2 & 0.71 & 500 \\ 
7 & 2 & 1.00 & 700 \\ 
8 & 2 & 1.43 & 1000 \\
\hline
\label{tab:expTPcondition}
\end{tabular}
\tablecomments{$\ast$ means the averaged temperature and its maximum and minimum of the eight monitoring points.}
\end{table}

\begin{figure}
\centering
\includegraphics[width=\linewidth]{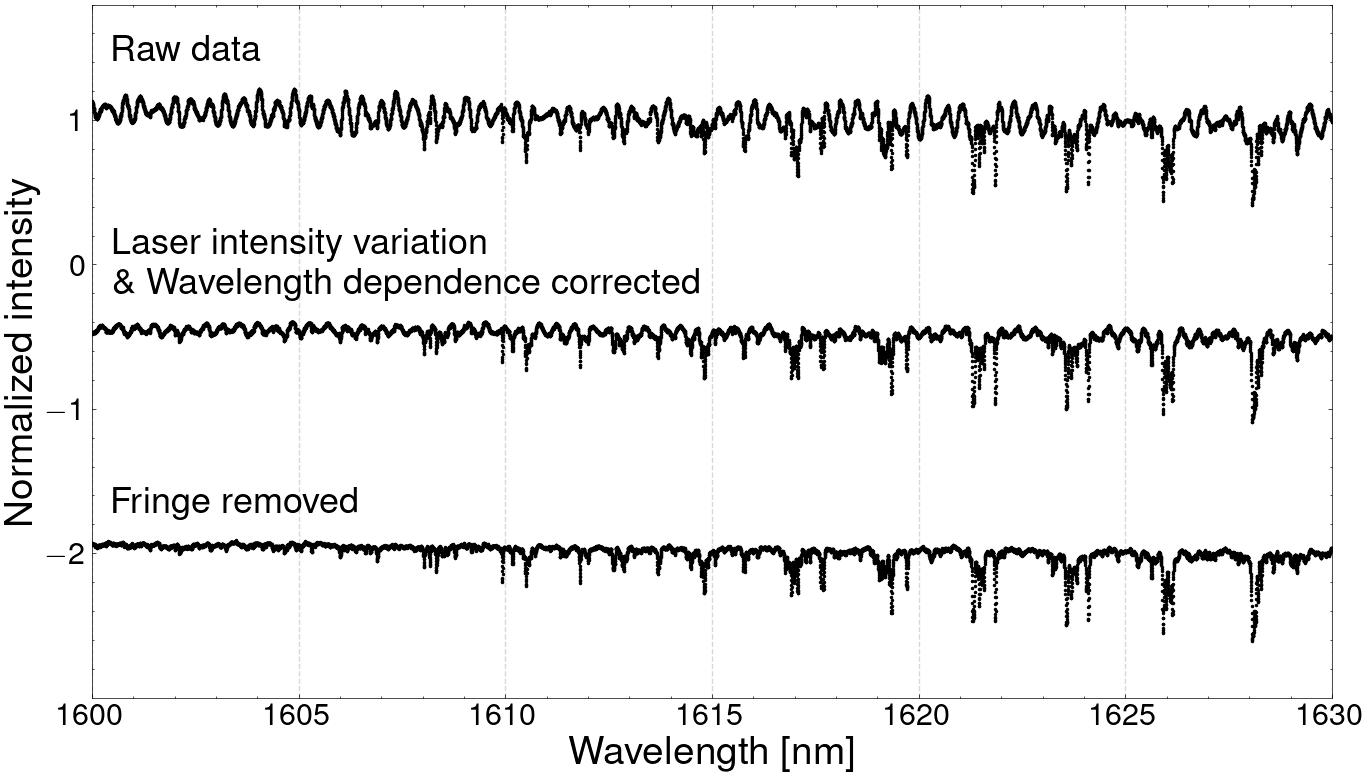}
\caption{CH\(_4\) 100\% spectrum following each data processing step. Top: Raw spectrum from the gas cell path; Center: Spectrum after correcting for laser intensity variation and wavelength dependence; Bottom: Spectrum after the fringe removal.}
\label{fig:CH41Spectra_DataReduction}
\end{figure}

To obtain accurate gas absorption spectra from the raw data, it is essential to correct for laser intensity fluctuations and the intensity trend across the wavelength caused by the optical components. The reduction process can be summarized as follows: First, we calibrate the spectra by dividing the gas cell raw spectra by the reference spectra from the simultaneous intensity monitoring, which effectively removes variations in laser intensity. Next, to account for the wavelength dependence of the spectrum, the intensity-corrected gas cell spectra are divided by the null spectrum. In the 100\% CH\(_4\) spectra, after applying these calibration procedures, we found a residual periodic intensity variation along the wavelength. It arises from the optical interference of transmitted light within the gas cell, occurring at both surfaces of the gas cell windows, which consist of a plane-parallel glass plate. To eliminate this interference pattern from the gas cell spectra, we fitted a fringe model, as described in Appendix~\ref{Frindge_removal}, along with the absorption model from HITEMP and a cubic polynomial function. The final gas cell spectra were derived by dividing the fitted fringe model and the polynomial component. The spectra for each reduction step of 100\% CH\(_4\) spectra are shown in Figure~\ref{fig:CH41Spectra_DataReduction}.
 In contrast, in the 10\% CH\(_4\) spectra, such interference patterns are absent thanks to the wedged windows of the gas cell.

\subsection{Results}
\begin{figure*}
\centering
\includegraphics[width=0.49 \linewidth]{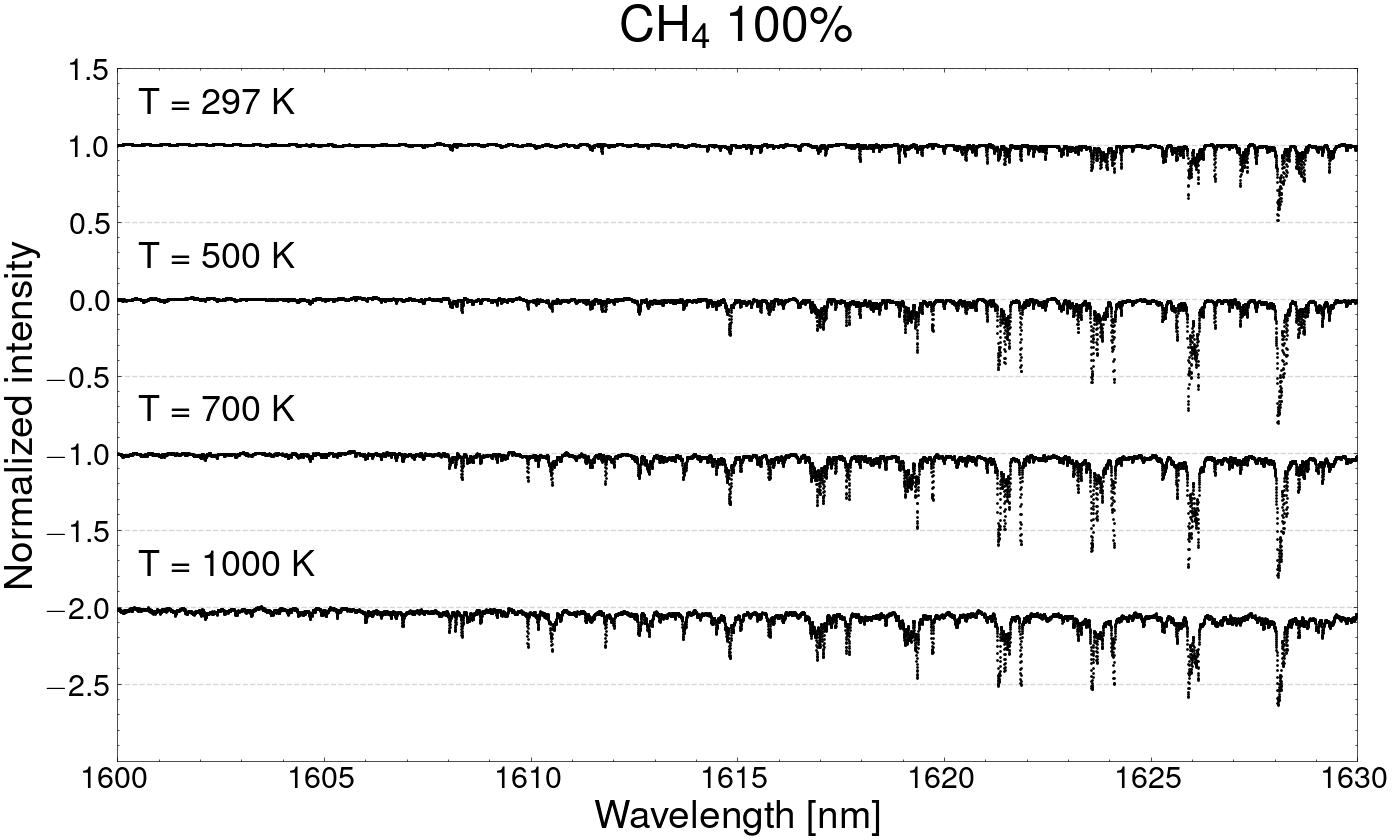}
\includegraphics[width=0.49 \linewidth]{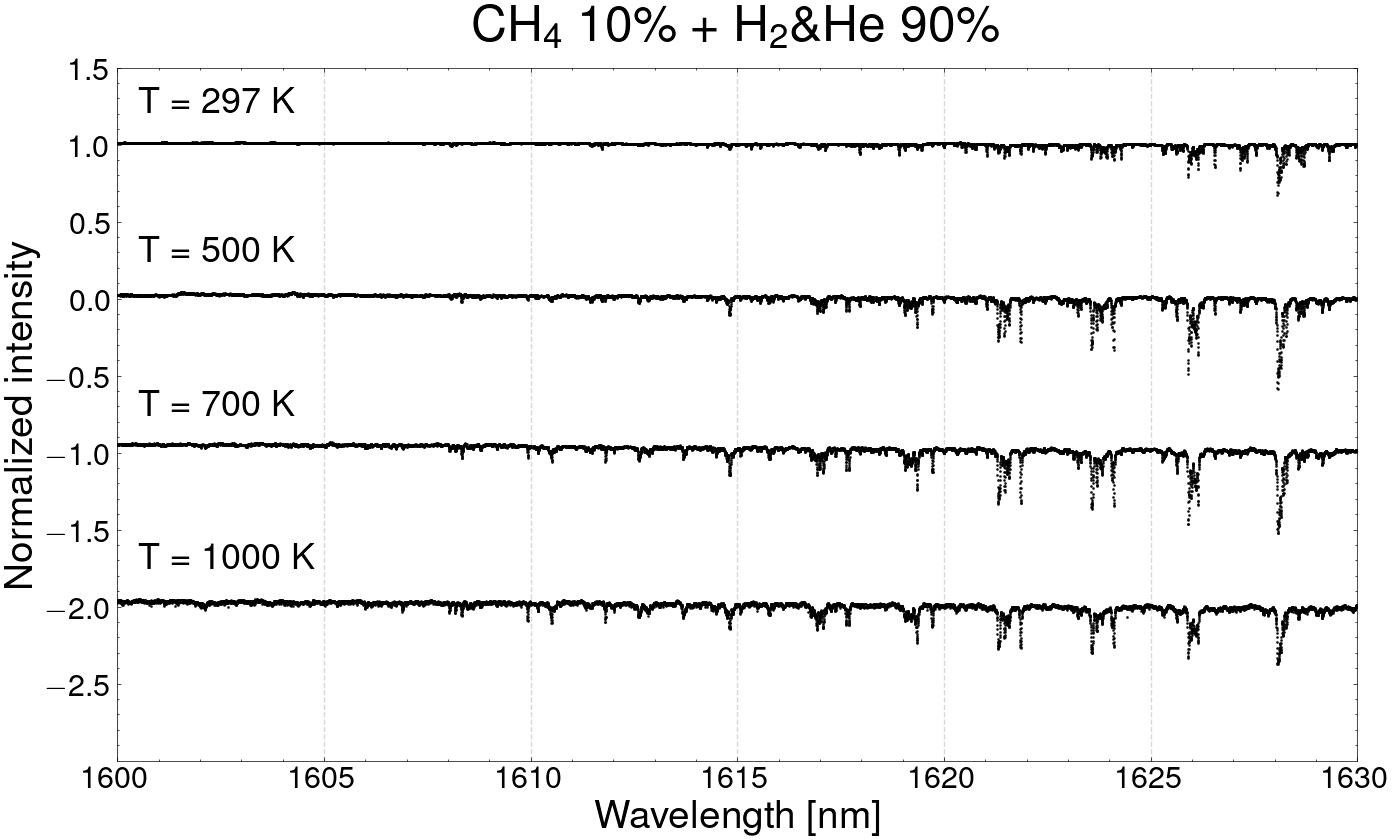}
\caption{CH\textsubscript{4} 100\% spectra (left) and 10\% (right) recorded at various temperatures. All spectra have been normalized based on the mean intensity across the entire wavelength range, with each spectrum offset by -1. }
\label{fig:CH4-AllSpectra_300-1000K}
\end{figure*}

Figure~\ref{fig:CH4-AllSpectra_300-1000K} show
the processed final spectra at 4 different temperatures in $\lambda$ = 1600--1630~nm, 10\% CH\(_4\) and 100\% CH\(_4\) respectively. The signal-to-noise (S/N) ratio for individual lines ranges from 15 to 150 derived from the inference results.\par

We verified the consistency of the overall line position by comparing the measured CH\(_4\) 100\% spectrum with another experiment and theoretical models. One of the previous studies, \cite{wong2019AtlasExperimentalTheoretical} determined the cross section for the CH\(_4\) 100\% spectra within the wavelength range of 1080--1920~nm at temperatures ranging from 295 to 1000~K and at a pressure of 0.133 bar. We transform this cross-section into transmittance, aligning the number density to match this study. While the absorption line depth and width are different due to different pressures and spectral resolutions, the positions of the lines and the intensity ratio of neighboring lines align with those reported in \cite{wong2019AtlasExperimentalTheoretical}.

\begin{figure*}
\centering
\includegraphics[height=0.9\textheight, keepaspectratio]{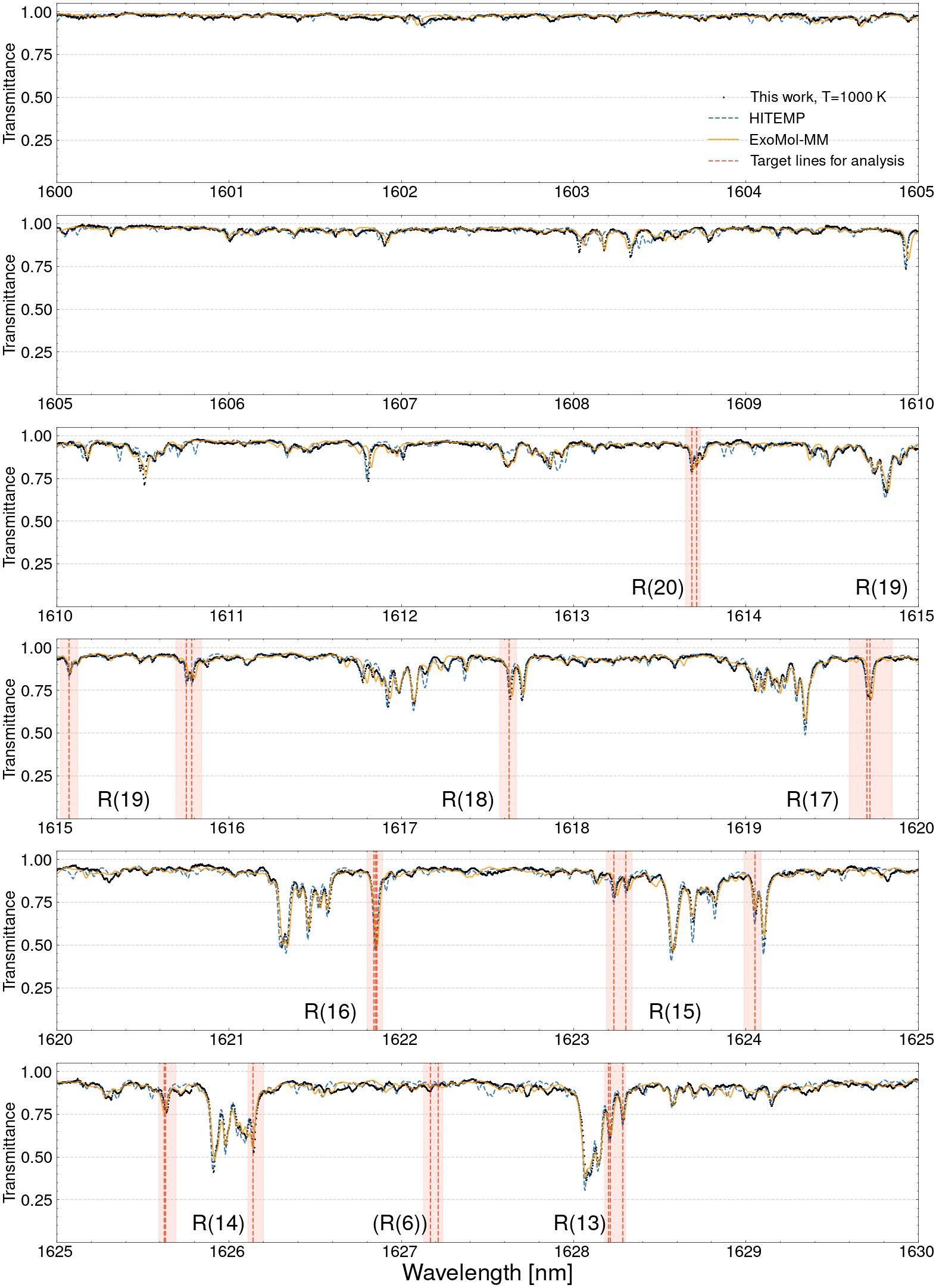}
\caption{
Comparison of the measured 100\% CH\(_4\) spectrum at \( T = 1000 \)~K and \( P = 1.43 \) bar (black) with model spectra using the HITEMP (dashed blue) and ExoMol (solid orange) databases. For the measured spectrum, the relative intensity was converted to transmittance by adjusting the intensity scale at low-absorption wavelengths to align with the transmittance of the HITEMP model. The colored area and red vertical lines indicate the 12 wavelength ranges and 22 target lines used to infer broadening parameters in Section \ref{sec:modeling}. R(13)--R(20) represent the region corresponding to the \( J_\text{lower} \) lines in the R-branch of the \( 2 \nu_3 \) band, while R(6) denotes the \( J_\text{lower} = 6 \) line in the R-branch of the unknown vibration mode.}
\label{fig:speccompare_HITEMPExomol}
\end{figure*}

Figure~\ref{fig:speccompare_HITEMPExomol} illustrates a comparison between the modeled spectra derived from the two main databases, HITEMP and ExoMol-MM \citep{tennyson_2024_2024,yurchenko_exomol_2024}, for spectra at T= 1000~K. Both models assume the same temperature, path length, and pressure as experimental conditions. Both databases well model most of the spectra; however, some discrepancies appear within specific wavelength ranges, such as $\lambda$ = 1608--1609, 1610--1611, and around 1617~nm. We believe that this discrepancy arises because the HITEMP database is supplemented \textit{ab initio} theoretical models for lines where some experimental data are missing~\citep{hargreaves_accurate_2020}.

\section{Bayesian Modeling of Line Profile}\label{sec:modeling}
% KM edit 11/2; previous version is in section3_20241101.tex

\newcommand{\wdop}{\alpha_D}
\newcommand{\namb}{n_{\mathrm{H_2\&He},l}}
\newcommand{\nself}{n_{\mathrm{self},l}}
\newcommand{\nhhe}{\namb}%${n_{\mathrm{H_2\&He},l}}
\newcommand{\gammahhe}{\gamma_{\mathrm{H_2\&He,ref},l}}
\newcommand{\gammaamb}{\gammahhe}%{\gamma_{\mathrm{amb,ref},l}}
\newcommand{\gammaself}{\gamma_{\mathrm{self,ref},l}}
\newcommand{\nuoffset}{\Delta\nu_\mathrm{offset}}
\newcommand{\pself}{P_\mathrm{self}}
\newcommand{\supk}{{(k)}}
\newcommand{\supkl}{{(k\Lambda)}}
\newcommand{\supl}{{(\Lambda)}}

\begin{figure*}[t]
\centering
\includegraphics[width=0.49\linewidth]{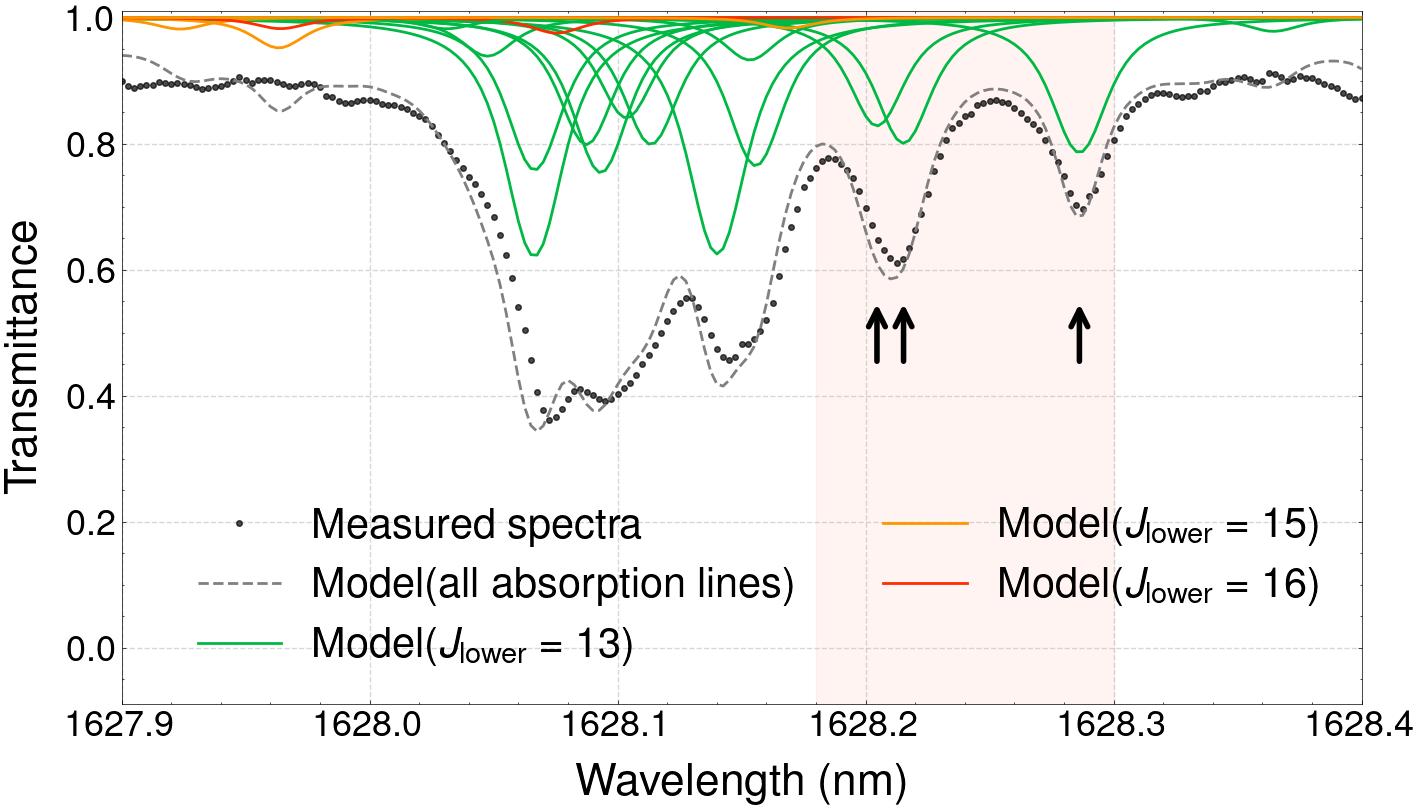}
\includegraphics[width=0.49\linewidth]{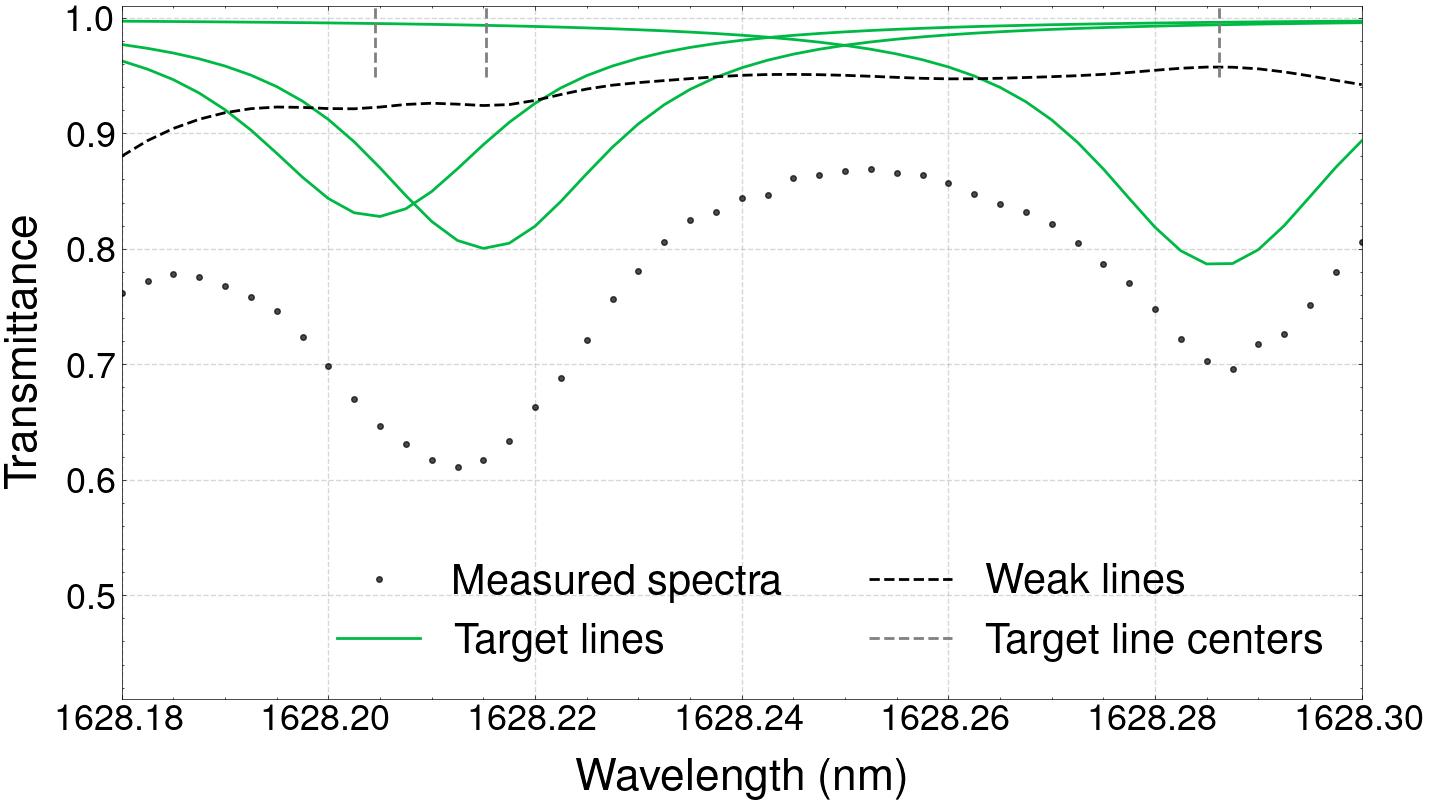}
\caption{Left: Comparison between the measured 100\% CH\(_4\) spectrum (black dots) and the theoretical model using HITEMP in the line-blended region of \(\lambda = 1627.9 - 1628.4\) nm at \( T = 1000 \)~K. In the theoretical model, the colored lines represent the contributions from 20 strong transitions, with colors identified by the corresponding \( J_\text{lower} \) values. The gray dashed line shows the model including all absorptions. The colored area indicates one of the wavelength ranges, and the positions of the target lines are marked by arrows. Right: The target lines (green) and weak lines (dashed line) in the spectral region of \(\lambda = 1628.18 - 1628.30\) nm correspond to the colored area in the left panel. The model for the weak lines is calculated by including all absorption lines except for the three target lines.}
\label{fig:line_example}
\end{figure*}

In this section, we model the spectra obtained in the eight different setups to determine the pressure-broadening parameters.
Several wavenumber regions of the spectra were selected to infer the broadening parameters of absorption lines. This selection was necessary because heavily blended lines appear in specific regions, making the analysis challenging. Additionally, including many lines in the inference results in the large number of free parameters and demands significant computational resources. We focus on the 22 strong, relatively well isolated lines shown by the red lines in Figure~\ref{fig:speccompare_HITEMPExomol}, as listed in Table~\ref{tab:inference_result_wav}, which we denote as ``target lines.''
The lower states of the rotational quantum numbers for these transitions are \( J_\mathrm{lower} = 13\text{--}20 \)  and \( J_\mathrm{lower} = 6 \). The vibrational modes of these lines are not listed in HITEMP; however, the former set aligns with the extrapolated positions of \( 2 \nu_3 \) for \( J_\mathrm{lower} = 13\text{--}20 \) \citep[e.g. Equation (1) in ][]{Nelson1948}, suggesting that they belong to the \( 2 \nu_3 \) band. 
The vibrational mode for the latter line with \( J_\mathrm{lower} = 6 \) is unknown, but likely  \(2\nu_2+\nu_3\) from the positions of the other known transitions (R(0), R(3), and R(6)) of this vibration mode in the HITEMP database.
All of these lines belong to the R-branch, meaning they are transitions where \( J_{\text{upper}} - J_{\text{lower}} = 1 \).
The left panel of Figure~\ref{fig:line_example} shows examples of target lines and strong non-target lines. Lines with numerous blended components, even among the strong lines, were excluded from this analysis due to the difficulty in accurately estimating their width.
Each setup (100\% vs 10\%  CH\(_4\), four different temperatures, labeled by $k=1,2,\dots,8$, as listed in Table \ref{tab:expTPcondition}) is characterized by the temperature $T^\supk$ and the partial pressure of  CH\(_4\) $\pself^\supk$ relative to the total pressure $P^\supk$ determined by $T^\supk$. Thus, by modeling these lines in the eight setups %with different temperatures and CH4 mixing ratios 
simultaneously, we can derive the temperature dependence of the broadening parameters, while also isolating the contributions from collisions with the ambient H\(_2\)/He gas molecules and the ``self'' collisions between  CH\(_4\) molecules.
Because these target lines are often blended with weaker lines, we also model the other ``weak lines'' simultaneously with the target lines. The right panel of Figure~\ref{fig:line_example} shows an example of the contribution of weak lines and a target line.
The weak lines are usually shallow and/or heavily blended and provide a limited amount of information about their line shapes,
%limited (or they just look like a continuum), 
so we do not fit for their line-shape parameters but model them as a fixed background.
We model the spectral lines using {\sf ExoJAX} \citep{kawahara_autodifferentiable_2022, 2024arXiv241006900K}, a differentiable spectral model built with {\sf JAX} --- an automatic differentiation and accelerated linear algebra package in Python. We leverage these capabilities to efficiently infer a large number ($\sim 100$) of parameters involved in our model.

\subsection{Line Profile Model}

The cross-section for a single line labeled by $l$ is modeled as:
%associated with the transition between a lower state $i$ and an upper state $j$ is expressed as:
\begin{equation}
\label{eq:cross_section}
%\sigma_V(\nu; \nu_{ij}, P,T) = S_{ij}(T) f_V(\nu; \nu_{ij}, P,T), 
\sigma_V(\nu; \nu_l, P,\pself,T) = S_l(T) f_V(\nu; \nu_l, P,\pself,T), 
\end{equation}
where $\nu_l$ is the wavenumber of the line center, $S_l(T)$ is the line strength that depends on the temperature $T$ and the energies of the lower and upper states associated with the transition, and
$f_V(\nu; \nu_l, P,\pself,T)$ is the line profile. 
The line strength $S_l$ can be computed using the molecular databases, as described in Section 2.2 in \cite{kawahara_autodifferentiable_2022}. For the line profile $f_V$, we adopt the Voigt profile given as the convolution of  Gaussian and Lorentzian profiles:
\begin{align}
\label{eq:Voigt_function}
&f_V(\nu; \nu_l, P,\pself,T) = \nonumber \\
&\int_{-\infty}^{\infty}   f_G(\nu - \nu^\prime; \nu_l, T) f_L(\nu'; \nu_l, P,\pself,T)  
d\nu^\prime.
\end{align}
Here, the Gaussian profile
\begin{equation}
f_G(\nu; \nu_l, T) = \sqrt{\frac{\ln 2}{\pi \alpha_D^2}} \exp\left( - \frac{(\nu - \nu_l)^2 \ln 2}{\alpha_D^2} \right)
\end{equation}
describes the Doppler broadening due to thermal motion. Its half width at half maximum (HWHM) width $\wdop(T)$ ~(cm$^{-1}$)is given by:
\begin{equation}
\label{eq:dopplar_width}
\wdop(T) = \frac{\nu_l}{c} \sqrt{\frac{2N_A k_{\rm B} T \ln 2}{M}},
\end{equation}
where \(M\) is the molecular mass~(g mol$^{-1}$), $k_{\rm B}$ is the Boltzmann constant, and \(N_A\) is the Avogadro constant. 
The Lorentzian profile
\begin{align}
\notag
f_L(\nu; \nu_l, &P,\pself,T) \\
&= \frac{1}{\pi} \frac{\gamma_l(P,\pself, T)}{\gamma_l(P,\pself,T)^2 + (\nu - \nu_l)^2},
\end{align}
describes the pressure broadening due to collisions with ambient molecules (H$_2$ and He in our case) as well as the ``self'' collisions between  CH\(_4\) molecules. 
In high-pressure atmospheres as considered here, the natural broadening (whose profile is also given by a Lorentzian) is negligibly small compared to the pressure broadening and is not included in our model.

The Lorentzian HWHM width $\gamma_l(P,\pself,T)$ (cm$^{-1}$/atm) is the parameter of our main interest in this work. Its dependence on the temperature and mixing ratio (partial pressure of  CH\(_4\)) is modeled following the formulation in the HITRAN/HITEMP database:
\begin{align}
\notag
\gamma_l(P,\pself,T) 
\label{eq:Lorentz_width_HITEMP}
&= \left( \frac{T_\text{ref}}{T} \right)^{\namb}  \gammaamb (P - \pself) \\
&\quad + \left( \frac{T_\text{ref}}{T} \right)^{\nself} \gammaself \pself . 
\end{align}
Here $P_{\rm self}$ is the partial pressure of the  CH\(_4\) gas: $P_{\rm self}=P$ for  CH\(_4\) 100\% and $P_{\rm self}=0.1P$ for  CH\(_4\) 10\%.
Modeling the temperature dependence is more suitable when the double power law is applied, especially over a wide temperature range such as this work~\citep{gamache_temperature_2018,yousefi_line_2021}. In this analysis, the single power law was adopted based on the HITEMP definition, since the double power law could not be applied with good precision due to the limited temperature conditions and the S/N of absorption lines.
The broadening due to collisions with the ambient molecules (i.e., H$_2$ and He) and self-collisions between  CH\(_4\) molecules are described by $\gammaamb$ and $\gammaself$, respectively, which are the values evaluated at the reference pressure $P_\mathrm{ref}=1\,\mathrm{atm}$ and temperature $T_\mathrm{ref}=296\,\mathrm{K}$:
\begin{align}
\gammaamb &\equiv \gamma_{\mathrm{H_2\&He},l}(P_\text{ref}, T_\text{ref}) \\
\gammaself &\equiv \gamma_{\mathrm{self},l}(P_\text{ref}, T_\text{ref})
\end{align}
The exponents $\namb$ and $\nself$ denote the temperature dependence of the broadening width of each line due to collisions with H\(_2\)/He and CH4 molecules, respectively.

\subsection{Transmittance Model for a Single Gas-cell Setup}

For a spectrum obtained from each setup labeled by $k$ ($k=1,2,\dots,8$), the total cross section $\sigma^\supk(\nu)$ is given as the sum of contributions from the target and weak lines:
\begin{align}
\notag
    \sigma^\supk(\nu; &P^\supk, T^\supk, \pself^\supk) \\
\label{eq:cross_section_total}
    &=  \sigma^\supk_{\rm targ}(\nu; P^\supk, T^\supk, \pself^\supk) \\
\notag
    &+  \sigma^\supk_{\rm weak}(\nu; P^\supk, T^\supk, \pself^\supk).
\end{align}

The cross-section for the target lines $\sigma^{(k)}_{\rm targ}(\nu)$ %in the $k$th sepctrum 
is modeled as the sum of Equation~\ref{eq:cross_section} for the 
%all the 22 
target lines labeled by $l$: 
\begin{align}
\notag
    &\sigma^\supk_{\rm targ}(\nu; P^\supk, T^\supk, \pself^\supk)\\
\label{eq:sigma_target}
    &= \sum_{l\in\mathrm{target}} \alpha_l \sigma_V(\nu; \nu_l, P^\supk, \pself^\supk, T^\supk)
\end{align}
The line center $\nu_l$ and the line strength $S_l(T_{\rm ref})$ in $\sigma_V$ (see Equation~\ref{eq:cross_section}) are fixed to the values taken from HITEMP. 
The parameters $\namb$,  $\nself$, $\gammaamb$, and $\gammaself$ in Equation~\ref{eq:Lorentz_width_HITEMP} are estimated for each line. 
The scale factor $\alpha_l$ is introduced to compensate for the error of the line strength $S_l(T_{\rm ref})$ adopted from HITEMP (whose uncertainty is estimated to be $\geq 20\%$). 
The cross sections of the target lines were calculated using {\sf OpaDirect}, one of the opacity calculators implemented in {\sf ExoJAX}; see Section 2.1 in \cite{kawahara_autodifferentiable_2022}.

Similarly, the cross-section for the weak lines (all the lines other than the target lines), $\sigma^{(k)}_{\rm weak}(\nu)$, in the $k$th spectrum is:
\begin{align}
\notag
%    \sigma^\supk_{\rm weak}(\nu)= \sum_{l\in\mathrm{weak}} \sigma_V(\nu; \nu_l+\nuoffset^\supk, P^\supk, \pself^\supk, T^\supk).
    &\sigma^\supk_{\rm weak}(\nu; P^\supk, T^\supk, \pself^\supk) \\
\label{eq:sigma_weak}
    &= \sum_{l\in\mathrm{weak}} \sigma_V(\nu; \nu_l, P^\supk, \pself^\supk, T^\supk).
\end{align}
As we did for the target lines, $\nu_l$ and $S_l$ in $\sigma_V$ are fixed to the HITEMP values. 
Unlike for the target lines, we do not fit for $\namb$, $\nself$, $\gammaamb$, and $\gammaself$ but fix them to the ``air'' values from HITEMP, and do not introduce the $\alpha_l$ parameter either. Thus, the model for the weak lines introduces no new free parameters.
The cross sections of the weak lines were calculated using another opacity calculator in {\sf ExoJAX}, {\sf PreMODIT}, which is specifically designed to efficiently compute the cross sections for a large number of lines \citep{2024arXiv241006900K}.

The cross-section in Equation~\ref{eq:cross_section_total} is used to compute the transmittance in the $k$th spectrum as a function of the wavenumber:
\begin{align}
\notag
    &\mathcal{T}^\supk(\nu) \\
    &= \exp\left(- \int n_V^\supk \sigma^\supk(\nu; P^\supk, T^\supk, \pself^\supk)\,\mathrm{d}L \right),
\end{align}
where $n_V^\supk$ is the volume number density of  CH\(_4\) molecules in the $k$th gas cell setup, and integration is performed along the beam's path length denoted by $L$. 
To account for the inhomogeneous temperature distribution in each gas cell, we evaluate this integral by dividing each gas cell into eight subcells of equal volumes. The temperature of each subcell was applied the average of the measured temperatures by the thermocouple in every 5 seconds during an measurement (approximately 38 minutes per measurement) for each points. For each subcell, we compute $n_V$ from applied $T$ assuming the ideal gas law and that the entire cell is isobaric, and then sum up the integrand $n_V\sigma\Delta L$ evaluated for each subcell.

%混んでる線の領域と推定波長域の図

\begin{table*}[htb]
\centering
\caption{Priors and Model Parameters}
\label{tab:HMCparameters}

\begin{tabular}{lccccc}
\hline\hline
Parameter & Symbol & Unit & Type & Lower or Mean & Upper \\ 
\hline
\multicolumn{5}{l}{\bf Parameters for Target lines} \\ 
Lorentzian width of \(\text{H}_2\&\text{He}\) broadening at ref. temp.
& $\gammahhe$ & $\mathrm{cm}^{-1}$/atm & Uniform & 0 & 0.2 \\ 
Lorentzian width of self-broadening at ref. temp. 
& $\gammaself$ & $\mathrm{cm}^{-1}$/atm & Uniform & 0 & 0.2 \\ 
Temperature exponent for \(\text{H}_2\&\text{He}\) broadening & $\nhhe$ & - & Uniform & -2$^{\ast}$ & 2 \\ 
Temperature exponent for self-broadening & $\nself$ & - & Uniform & -2$^{\ast}$ & 2 \\ 
Line strength coefficient & $\alpha_l$ & - & Uniform & 0 & 5 \\ 

\hline
\multicolumn{5}{l}{\bf Common parameters for target and weak lines} \\ 
Wavenumber offset & $\nuoffset^\supkl$ & $\mathrm{cm}^{-1}$ & Uniform & -0.05 & 0.05 \\ 

\hline
\multicolumn{5}{l}{\bf Polynomial parameters} \\ 
3rd coefficient & $a^\supkl$ & - & Uniform & -64 & 64 \\ 
2nd coefficient & $b^\supkl$ & - & Uniform & -16 & 16 \\ 
1st coefficient & $c^\supkl$ & - & Uniform & -4 & 4 \\ 
0th coefficient & $d^\supkl$ & - & Uniform & 0 & 2 \\ 

\hline
\multicolumn{5}{l}{\bf Noise Parameter} \\ 
standard deviation of the Gaussian noise & $\sigma_{\rm noise}^\supkl$ & - & Exponential & 1E-3 & - \\
\hline
\end{tabular}
\flushleft{Note --- The parameters with the subscript $l$ is defined for each of the 22 target lines. 
The other parameters with the superscript $(k\Lambda)$ are defined for each of the 12 wavenumber regions of the eight spectra. 
The reference temperature (ref. temp.) is $T_{\rm ref} = 296\,\mathrm{K}$. \\
$^\ast$For the line of \(\lambda\) = 1621.84132, 1621.85218, and 1621.85931 nm (all in the same wavenumber region and inferred at the same time), 0--2 were adopted as the prior except for $\nhhe$ of the 1621.85931 nm line so that to avoid the significant drop of inference precision.
}
\end{table*}

\subsection{Bayesian Inference of the Model Parameters}
We analyze the target lines in the eight spectra simultaneously; this is essential to determine $\namb$, $\nself$, $\gammahhe$, and $\gammaself$ for each line using the data obtained in gas cell setups with different $T$ and different $\pself$.
Rather than modeling all 22 target lines at once, we extract $12$ wavenumber regions labeled by $\Lambda$, each containing 1--3 target lines and with the widths of about 0.085~nm to 0.2~nm, and fit the eight spectra in each region separately.
The width of each region is chosen so that it contains at least one target line, and that it does not contain the line centers of other prominent weak lines. The minimal width of 0.085~nm is wide enough to cover the entire wings of the target lines, whose FWHM is $\sim 0.025\,\mathrm{nm}$ even for $T$=1000~K and is smaller at lower temperatures.

We model the spectrum in the wavenumber region $\Lambda$ from the $k$th gas cell setup, $\mu^\supkl(\nu)$, as
\begin{align}
\notag
\mu^\supkl&(\nu) =  \mathcal{T}^\supk(\nu-\nuoffset^\supkl)\\ 
\label{eq:model_single}
&\times (a^\supkl \tilde{\nu}^3 + b^\supkl \tilde{\nu}^2 + c^\supkl \tilde{\nu} + d^\supkl).
\end{align}
Here the $\nuoffset^\supkl$ parameter is introduced to account for the uncertainty in the wavenumber calibration in each part of the spectrum and is estimated together with the other parameters. 
The polynomial part is introduced to account for trends of instrumental origin, where $\tilde{\nu}=\nu-\nu_{\rm min}$ with $\nu_{\rm min}$ being the smallest wavenumber in each region, and its coefficients are also estimated with the other parameters.
Then we define the following likelihood function for the flux data in the wavenumber region $\Lambda$, assuming that the noise obeys the independent and identical normal distribution:
\begin{align}
\label{eq:likelihood}
\notag
\mathcal{L}^\supl = &\prod_{k=1}^{8} \frac{1}{\sqrt{2\pi [\sigma_\mathrm{noise}^\supkl]^2}}\\
&\times\exp\left( -{{\sum_{i} [y^\supkl_i- \mu^\supkl(\nu_i)]^2}\over {2[\sigma_\mathrm{noise}^\supkl]^2}}  \right),
\end{align}
where $y_i^\supkl$ is the $i$th measured flux in the $\Lambda$th wavenumber region of the $k$th spectrum, $\nu_i$ is the corresponding wavenumber, and $\sigma_{\rm noise}^\supkl$ is the standard deviation of the noise within the $\Lambda$th wavenumber region of the $k$th spectrum. We estimate $\sigma_{\rm noise}^\supkl$ along with the other model parameters for each $\Lambda$.
Thus in total, the model for each wavenumber region $\Lambda$ containing $N_{\rm targ}^\supl$ target lines includes 
$N_{\rm param}^\supl = 5N_{\rm targ}^\supl + +6N_{\rm setup} = 5N_{\rm targ}^\supl + 48$
free parameters with $N_{\rm setup}=8$: $(\namb,\nself, \gammahhe, \gammaself, \alpha_l)$ for each target line, and $(\nuoffset^\supkl, a^\supkl, b^\supkl, c^\supkl, d^\supkl, \sigma_{\rm noise}^\supkl)$ for each wavenumber region of each spectrum. 
This number $N_{\rm param}^\supl$ ranges from 53 ($N_{\rm targ}^\supl=1$) to 63 ($N_{\rm targ}^\supl=3$) for each wavenumber region, and sums up to $\sum_{\Lambda=1}^{12} N_{\rm param}^\supl = 5\times 22 + 48 \times 12=686$ 
for all the $22=\sum_\Lambda N_{\rm targ}^\supl$ target lines in all the 12 wavenumber regions.

Given the large number of free parameters that are also strongly correlated, we conduct full Bayesian modeling using Hamiltonian Monte Carlo \citep[HMC;][]{1987PhLB..195..216D,2011arXiv1111.4246H} to infer model parameters by sampling from their joint posterior probability distribution, rather than the least squares fitting as commonly employed in similar studies \citep[e.g.][]{gharib-nezhad_h2-induced_2019, sung_h2-pressure_2020, yousefi_line_2021}. The HMC is a type of Markov Chain Monte Carlo (MCMC) method characterized by a higher acceptance rate, lower correlation between samples, and a milder dependence of computational time on the number of parameters compared to the MCMC with random-walk proposals \citep[e.g.][]{2012arXiv1206.1901N, 2024arXiv241006978B}.
For each wavenumber region $\Lambda$, we sample from the joint posterior distribution for the set of parameters $\theta$:
\begin{align}
    p^\supl(\theta|y) \propto \mathcal{L}^\supl(\theta) \pi(\theta)
\end{align}
adopting the prior $\pi(\theta)$ that is assumed to be separable for each parameter as summarized in Table~\ref{tab:HMCparameters}. We use the No-U-Turn Sampler \citep{2011arXiv1111.4246H} as implemented in {\sf NumPyro} \citep{phan2019composable}.
We ran a single chain for 1,000 warm-up steps and for 2,000 sampling steps, after which the split Gelman--Rubin statistics $\hat{R}$ was below 1.1 for all parameters, indicating that the chain is well mixed \citep{BB13945229}. 
The results are summarized in Table~\ref{tab:inference_result_wav}. The calculations were performed on an NVIDIA A100 GPU (either 40 or 80 GB) operating in parallel, and took approximately 4--6 hours to complete for each wavelength region.

%\clearpage
\section{Results}\label{sec:result}
We carried out the Bayesian inference for 22 strong spectral lines as the target line, which were divided into 1 to 3 lines in 12 distinct wavelength regions. 

% KM: moved detail to Section 3
In Section \ref{ss:example}, we first introduce the results of the Bayesian analysis for a representative case: a single wavelength region containing three target lines. In Section \ref{ss:alltargets}, we present the analysis results for all 22 target lines.

\subsection{Analysis results for the wavelength region 1628.18--1628.30 nm}\label{ss:example}

\begin{figure*}
\centering
    \centering
    \includegraphics[width=0.48\linewidth]{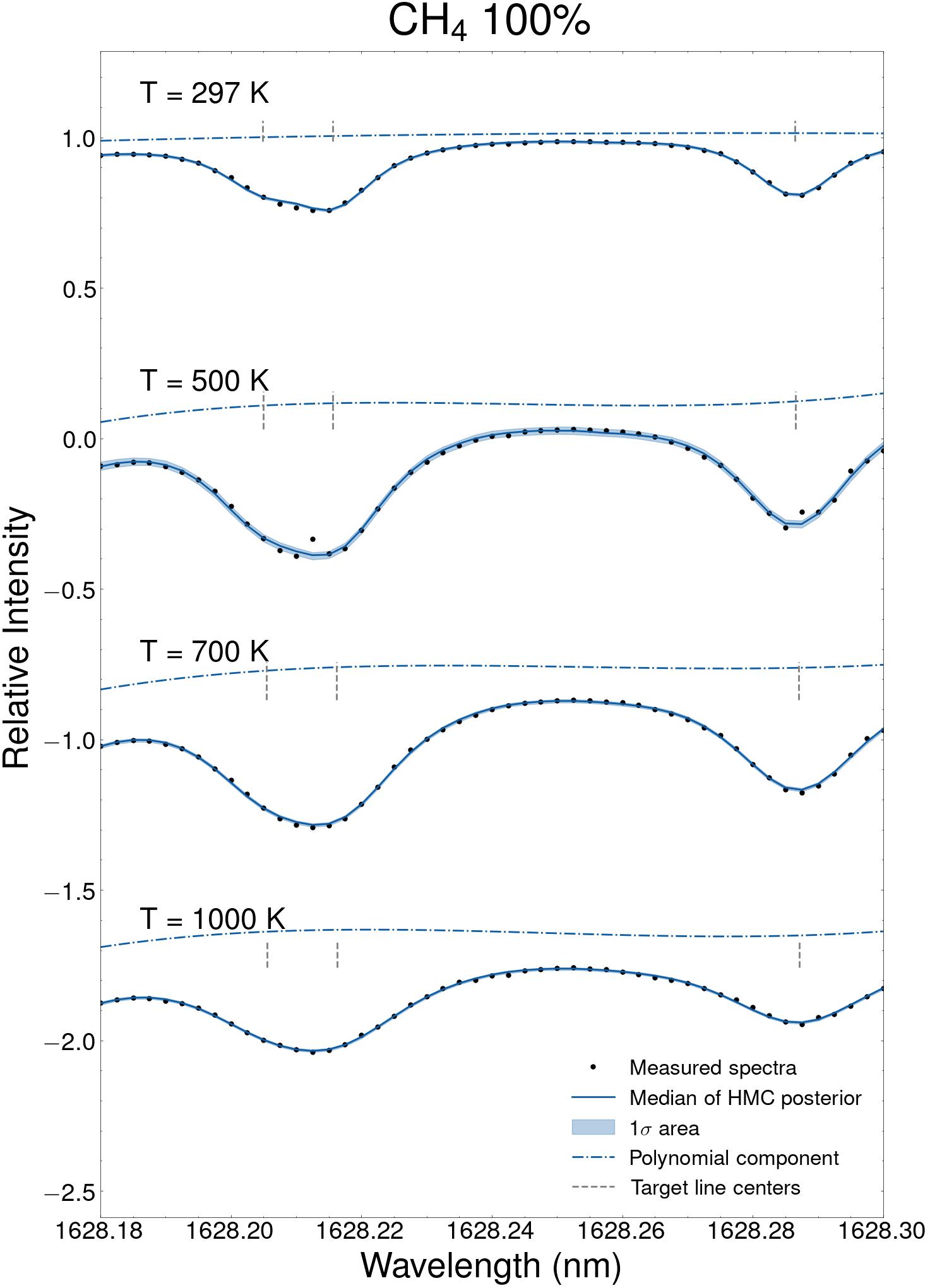}
    \includegraphics[width=0.48\linewidth]{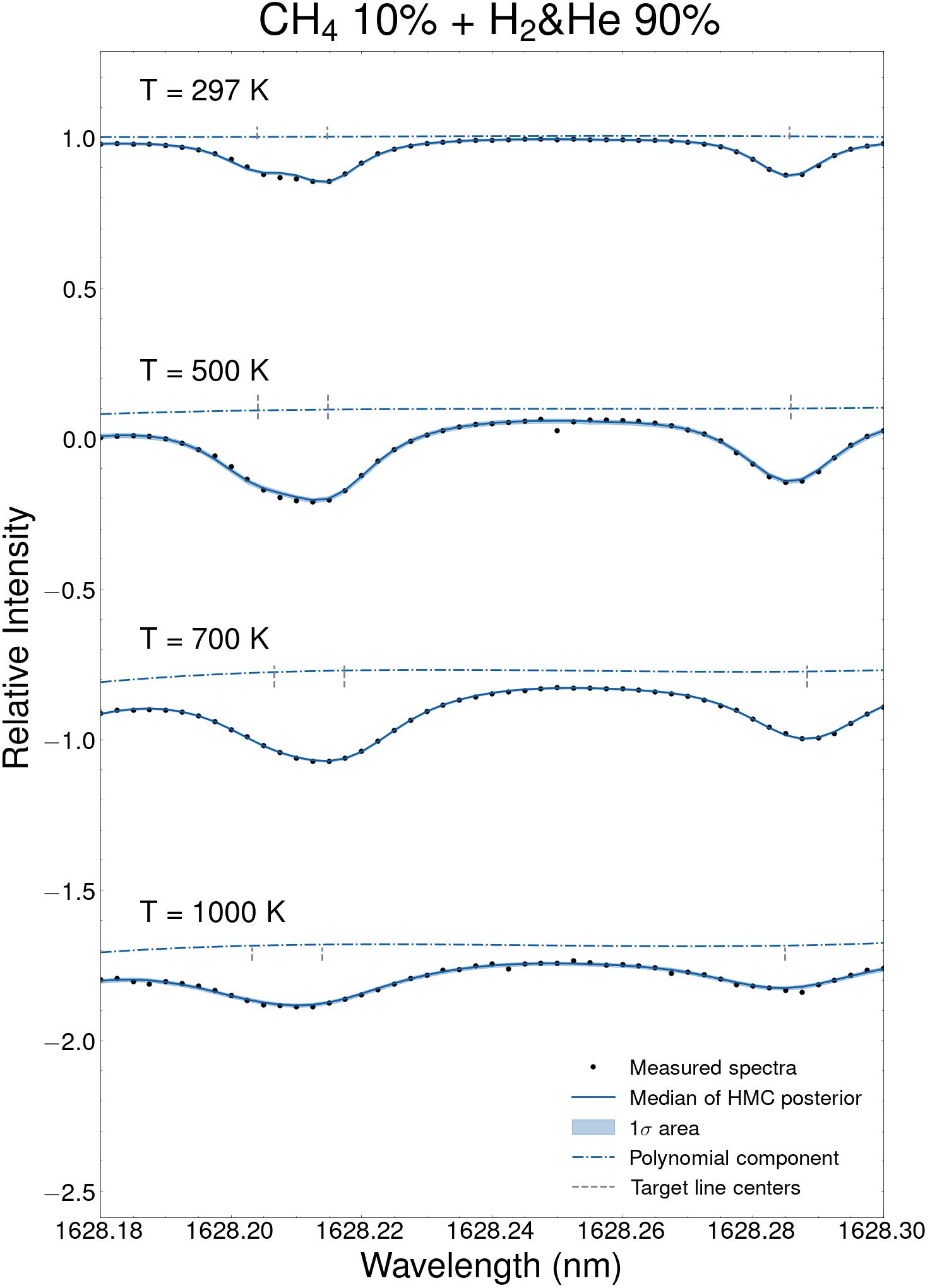}
\caption{Example of the eight spectra set for 100\% methane (left) and 10\% methane + 90\% H\(_2\)/He (right) at four temperatures (from top: 297, 500, 700, 1000~K) alongside predictions from HMC-NUTS, at $\lambda$ = 1628.18 - 1628.30 nm.  The measured spectra are shown by dots. The fit results appear as colored lines with median values from the posterior, and the colored regions signify 1$\sigma$ credible intervals. The blue dashed lines represent the polynomial components of each fitting result, while the gray dashed vertical lines indicate the position of the inferred three target line centers.}
\label{fig:HMCfit_162818}
\end{figure*}

Figure~\ref{fig:HMCfit_162818} illustrates an example of the eight spectra in a wavelength region 1628.18 - 1628.30 nm and their predictions inferred from the HMC-NUTS analysis. This wavelength region includes three target lines, $\lambda$ = 1628.28617, 1628.21526, and 1628.20455 nm. The measured spectra are depicted using dots, and the 1$\sigma$ credible intervals are shown by colored areas. The spectra on the left correspond to 100\% CH\(_4\), while those on the right show 10\% CH\(_4\), (the remaining 90\% is a mixture of H\(_2\&\)  He).  The spectra are organized vertically to represent varying temperatures $T$ = 297~K, 500~K, 700~K, and 1000~K from top to bottom. 
The model described in Section 3 successfully reproduces all eight measured spectra simultaneously, covering four different temperatures and methane concentrations of 10\% and 100\%.

%CH4 10%, 1000Kのスペクトルに関係するパラメーターのcorner plot
\begin{figure*}
\centering
\includegraphics[width=\linewidth]{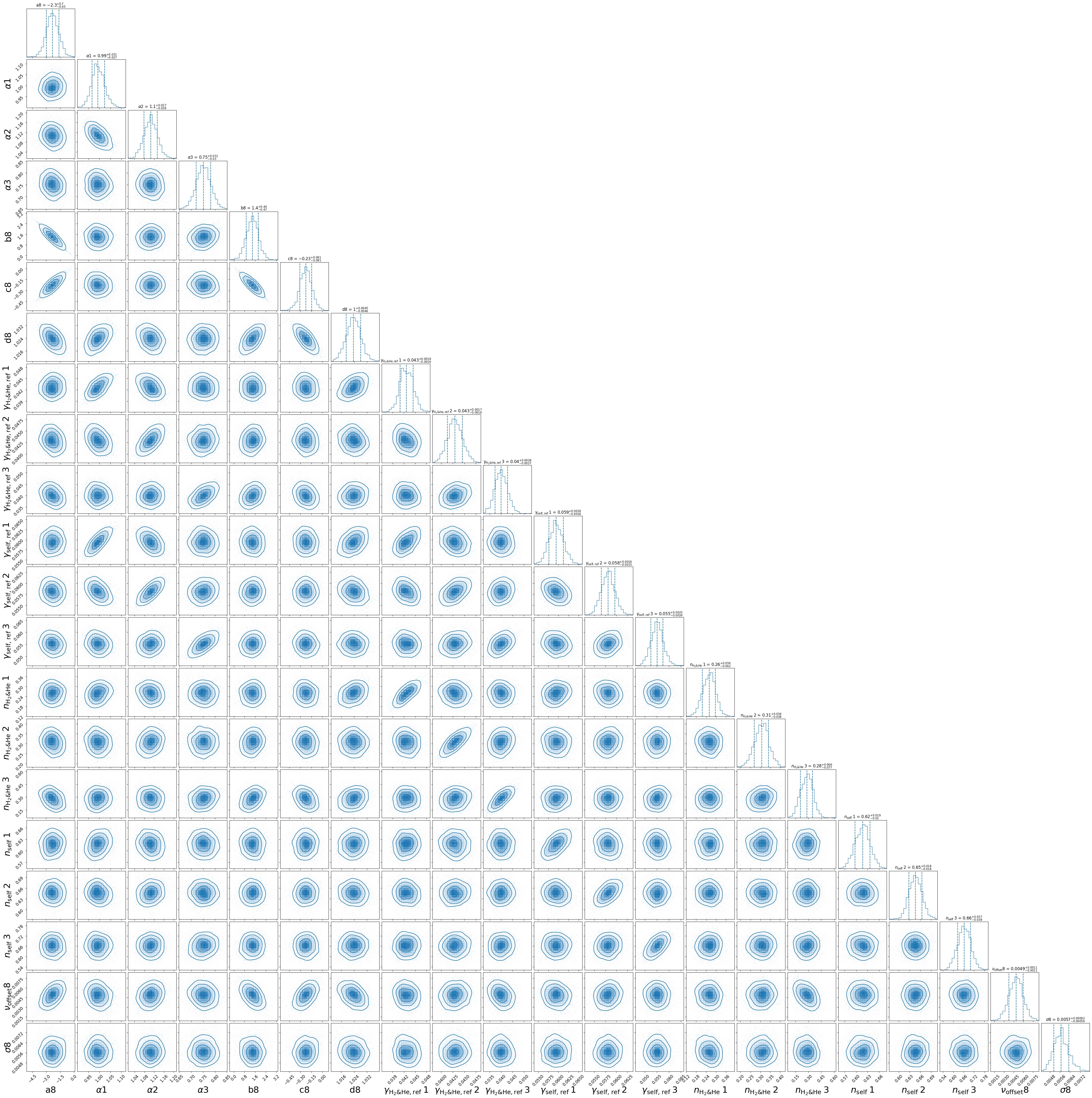}
\caption{Corner plots of $\lambda$ = 1628.18--1628.30 nm, only extracted the parameters related to the spectra of CH$_4$ 10\% at $T$ = 1000~K (measurement setup $k$ = 8) from a total 63 parameters. The suffixes of each broadening parameters ($\gamma$ and $n$) denote inferred lines ($\lambda$ = 1628.28617, 1628.21526, and 1628.20455 nm) in increasing wavenumber order (= inverse of wavelength).}
\label{fig:HMCcorner_162818_spec8}
\end{figure*}

Figure~\ref{fig:HMCcorner_162818_spec8} is a corner plot that shows only the parameters related to the 1000~K/10\% methane spectrum, extracted from the 63 parameters analyzed for the spectral range in Figure~\ref{fig:HMCfit_162818}.
Polynomial coefficients \(a, b, c, d\), line strength scaling factor \(\alpha\), and wavenumber offset \( \Delta \nu_\text{offset}\) exhibit positive or negative correlations, allowing other parameters to adjust and compensate for variations in absorption depth and position to fit the spectra accurately.
Weak positive correlations between \(\alpha\) and \(\gammaamb\) ( and \(\alpha\) and \(\gammaself\) ) are observed. This occurs likely because, assuming a constant line strength, an increase in \(\gamma\) leads to a shallow absorption depth, therefore, \(\alpha\) increases to maintain the line depth consistent with the spectra.

%γ, nのみのcorner plot
\begin{figure*}
\centering
\includegraphics[width=\linewidth]{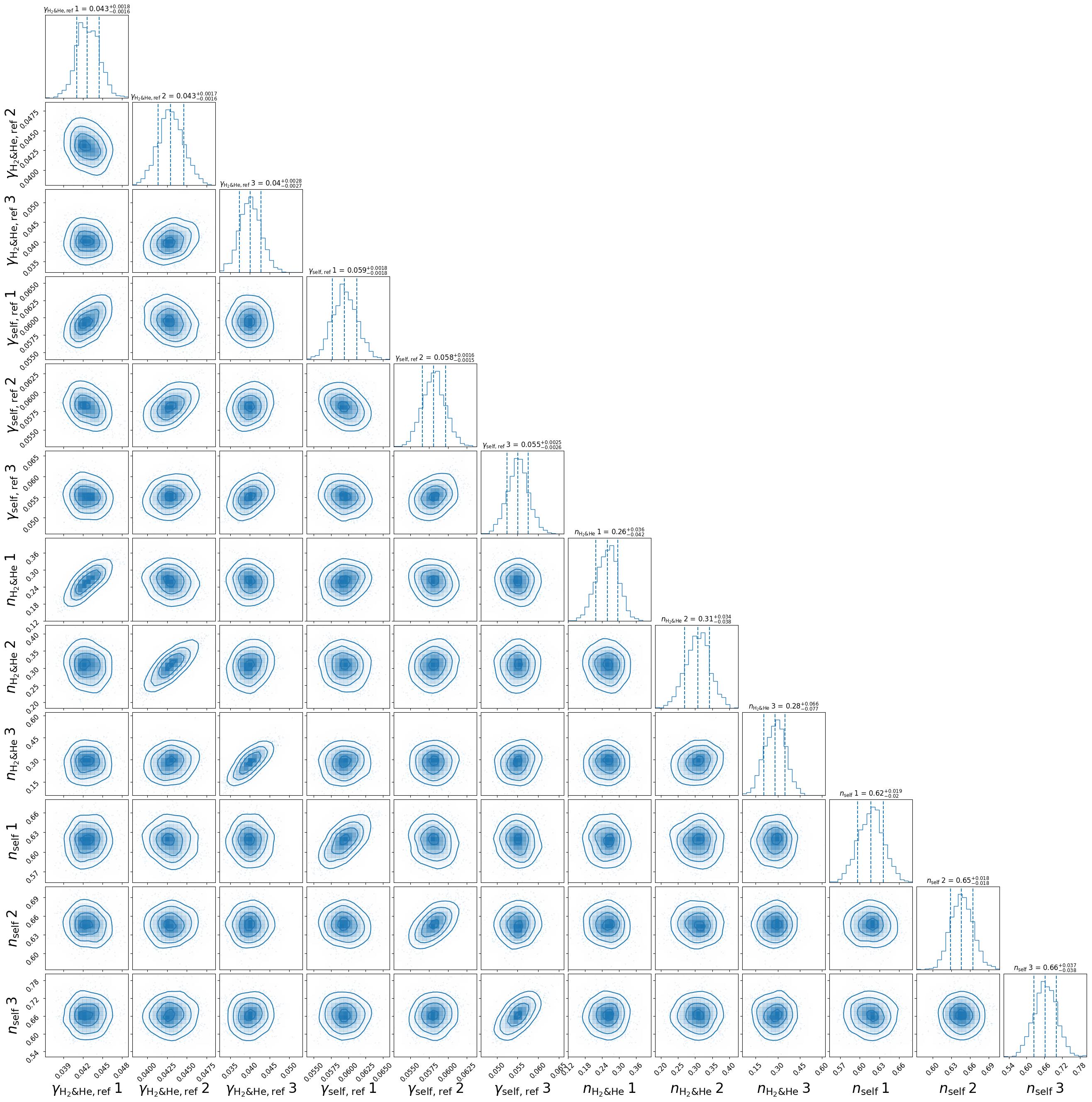}
\caption{Corner plot for the broadening parameters of the three target lines in $\lambda$ = 1628.18 - 1628.30 nm (same as Figure~\ref{fig:HMCcorner_162818_spec8}), selected from the total 63 parameters. The suffixes of each parameter denote inferred lines ($\lambda$ = 1628.28617, 1628.21526, and 1628.20455 nm) in increasing wavenumber order.}
\label{fig:HMCcorner_162818}
\end{figure*}

Figure~\ref{fig:HMCcorner_162818} displays corner plots that focus on the broadening parameters (\(\gamma_{\text{self},\text{ref}}\), \(\gamma_{\text{H}_2\&\text{He}, \text{ref}}\),  \(n_{\text{self},\text{ref}}\), and  \(n_{\text{H}_2\&\text{He}, \text{ref}}\)) for the three target lines within the wavelength range of Figure~\ref{fig:HMCfit_162818}. 
A positive correlation is observed between \(\gamma_{\text{self},\text{ref}}\) and \(n_{\text{self},\text{ref}}\), also \(\gamma_{\text{H}_2\&\text{He}, \text{ref}}\) and \(n_{\text{H}_2\&\text{He}, \text{ref}}\) for each target line. This is due to the relation where a larger \(n\) results in a smaller \(\gamma(P,T)\) at elevated temperatures, so each \(\gamma_{\text{ref}}\) increases to compensate for the effect at higher temperatures. 
A weak correlation also exists between \(\gamma_{\text{self},\text{ref}}\) and \(\gamma_{\text{H}_2\&\text{He}, \text{ref}}\). 
This weak correlation is interpreted as a pseudo-correlation, likely due to the strong correlations that \(\gamma_{\text{self},\text{ref}}\) and \(\gamma_{\text{H}_2\&\text{He}, \text{ref}}\) each has with \(n\).
This case demonstrates the importance of simultaneous inference to account for the influence of the self-broadening on the CH\textsubscript{4} 10\% spectra.

For the self-broadening parameters of these target lines, \(\gamma_{\text{self},\text{ref}} \) ranged from 0.055 to 0.059, \(n_{\text{self}}\) varied 0.62--0.66, both with the uncertainty in $\pm$ 10\%. 
In the HITEMP database, the three target lines have a common value of \(\gamma_{\text{self},\text{ref}} \) = 0.065 and \(n_{\text{air}}\) = 0.63, which are consistent with our measurement.

For the $\text{H}_2\&$He broadening of three target lines, we obtained 0.040--0.043 as \(\gamma_{\text{H}_2\&\text{He}, \text{ref}}\) with an error of about 10\% and \(n_{\text{H}_2\&\text{He}}\) varied 0.26--0.31, with the uncertainty of about 15--25\%.
ExoMol provides the broadening parameters for hydrogen-helium atmospheres of \(\gamma_{\text{H}_2\&\text{He}, \text{ref}} = \) 0.060 for these three lines \footnote{The value is calculated by summing \(\gamma_{\text{H}_2, \text{ref}}\) and \(\gamma_{\text{He}, \text{ref}}\) after multiplying by the volume mixing ratio \(\text{H}_2\):He = 84:16. The difference between \(\alpha_{\text{ref}}\) and \(\gamma_{\text{ref}}\) is adjusted by dividing by 1.01325 (1 atm in bar). }. 
Compared to the ExoMol, the  \(\gamma_{\text{H}_2\&\text{He}, \text{ref}}\) for those inferred lines are 30\% smaller, and the temperature exponent is milder.

The broadening data for \(\text{H}_2\) and He in Exomol are created independently from line list, uses the mean value for each quantum number, which are from several different experiments \citep[marked by $\ast$ in Table\ref{tab:previous_exp};][]{barton_exomol_2017}. These experiments cover lines up to \(J_{\text{lower}} \le 16\) in $\lambda$ = 3--7~{\textmu}m and in $T$ = 130--296~K. Therefore, the values for the target lines are not directly from experiments in the same condition as ours. It is not unnatural that our measured values differ from those of ExoMol. 

As a reference, the HITEMP database provides the air broadening parameter as  \(\gamma_{\text{air,ref}} \) = 0.049 with uncertainty $\ge$ 20\%.
Thus, the measured \(\gamma_{\text{H}_2\&\text{He}, \text{ref}}\) is also lower than \(\gamma_{\text{air,ref}}\) in HITEMP at $T$ = 296~K and $P$ = 1~atm. 

\begin{figure}[h]
\centering
\includegraphics[width=\linewidth]{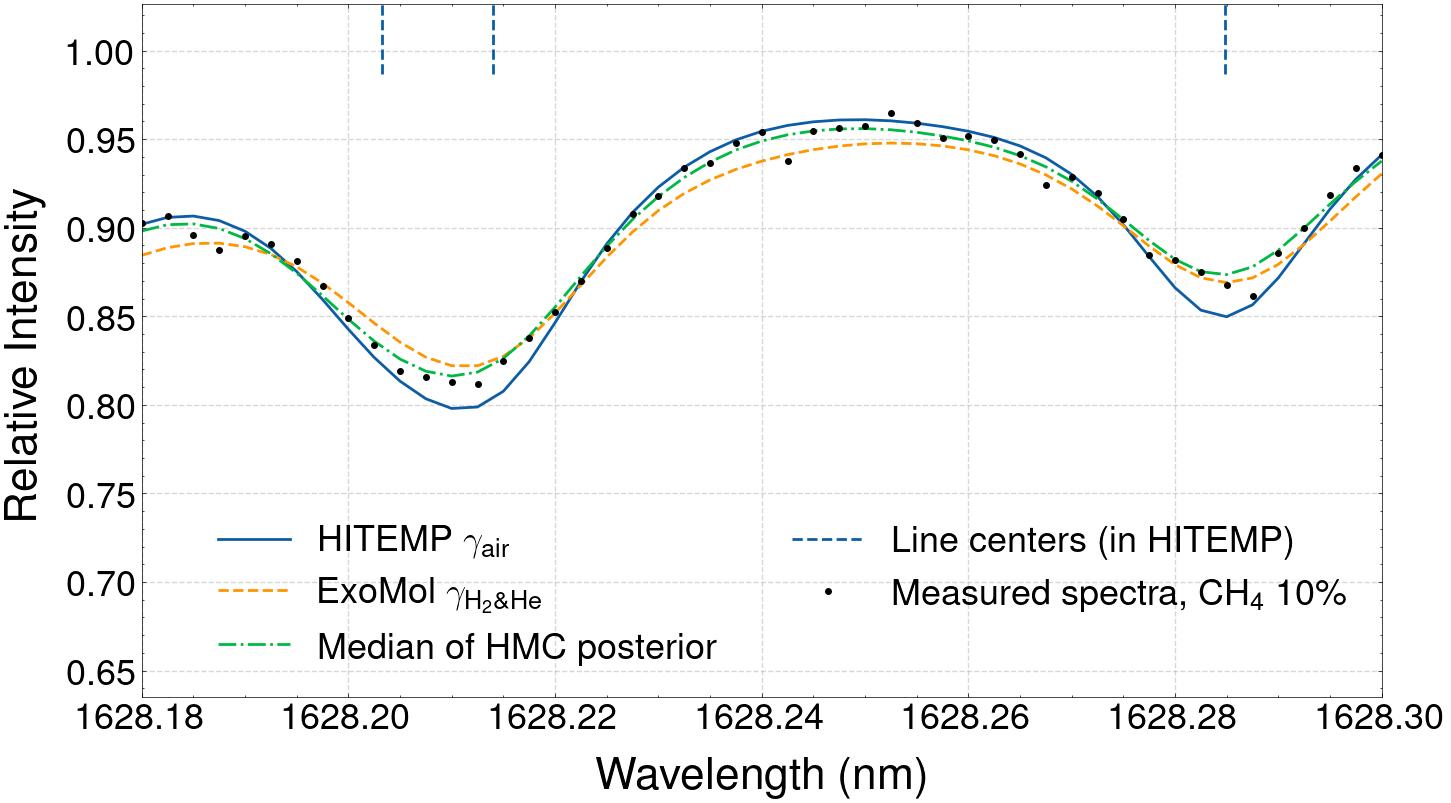}
\caption{Comparison of the measured spectra ($\lambda$ = 1628.18-1628.30 nm) with the modeled spectra. Black points: recorded spectra at \(T = 1000\,\text{K}\), Blue: using HITEMP \(\gamma_{\text{air}}\), Orange: utilizing ExoMol \(\gamma_{\text{H}_2} \) and \(\gamma_\text{He}\) along with the mixing ratio of the experiment, Green: using the median values of inference results. Note that for HITEMP and ExoMol models, the effect of \(\gamma_{\text{self}}\) was accounted by using the value of \(\gamma_{\text{self},\text{ref}}\) in HITEMP and \(\gamma_{\text{default}}\) for ExoMol.}

\label{fig:SpecCompare_162818}
\end{figure}

Figure~\ref{fig:SpecCompare_162818} shows the differences between the measured spectrum and the modeled spectra, which were generated using the broadening parameters obtained from the inference results, as well as those provided by ExoMol ($\text{H}_2\&\text{He}$) and HITEMP (air) for these three lines. As shown in this figure, neither the ExoMol $\text{H}_2$/He broadening parameters nor the HITEMP air broadening parameters can reproduce the experimental values for the three lines, in contrast to the well-replicated spectra by the inference results.
This fact suggests that current databases are insufficient for discussing line profiles in hydrogen-helium atmospheres of substellar objects, highlighting the importance of experimentally determined broadening parameters.

\subsection{Inference of the 22 target lines}\label{ss:alltargets}
%αの比較
\begin{figure}
\centering
\includegraphics[width=\linewidth]{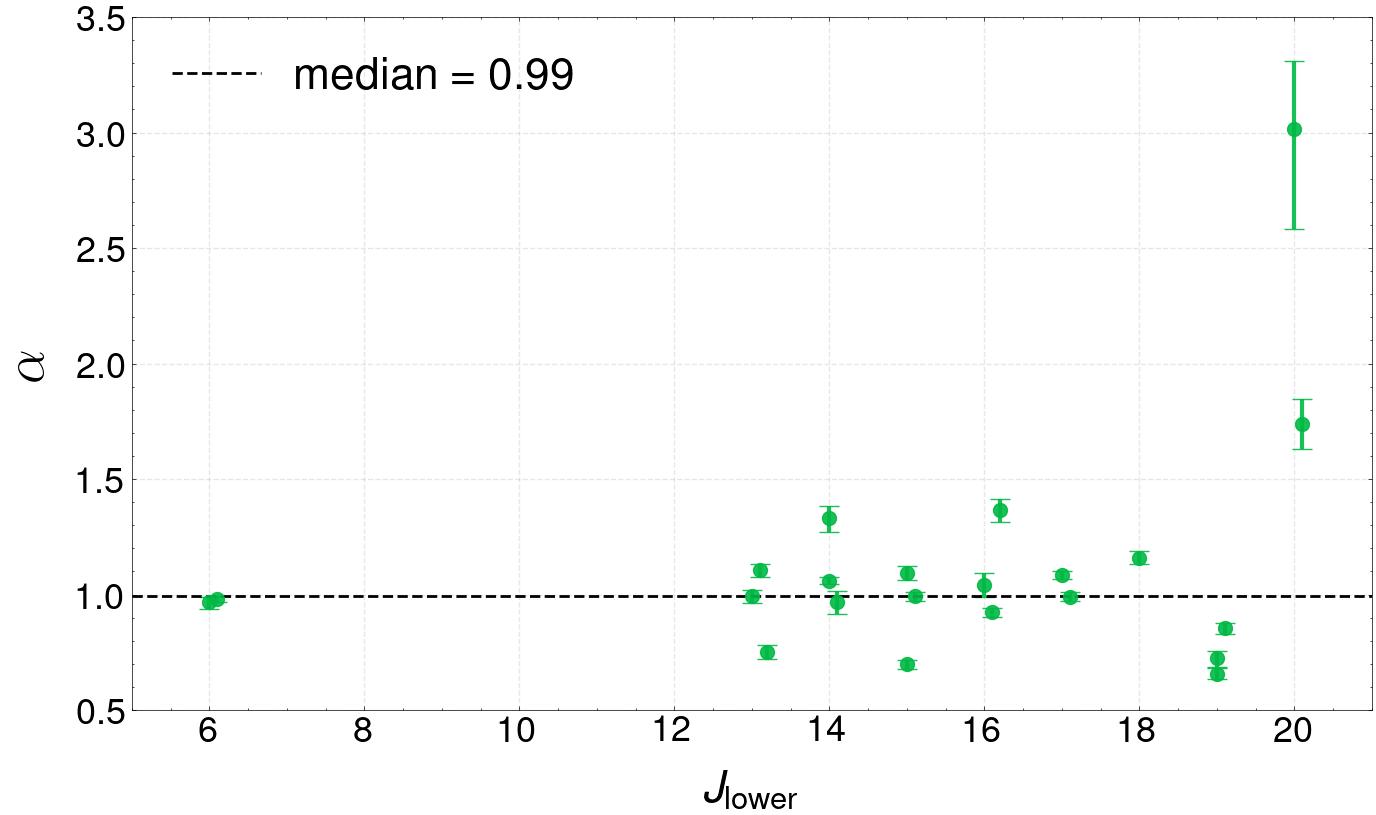}
\caption{Inference results of line strength scale factor $\alpha$ with quantum numbers. The markers with error bars represent the median values and the 1$\sigma$ credible intervals. For the ease of distinguishing each plots, there are offset if multiple lines correspond to the same \(J_{\text{lower}}\).}
\label{fig:alpha_Results_quantumnum}
\end{figure}

Following the same analysis approach as in Section \ref{ss:example}, we conducted analyses on 12 wavelength regions to estimate the broadening parameters for 22 target lines. These estimates are summarized in Table \ref{tab:inference_result_wav}.
It also summarizes the scale factor \(\alpha\) for line strength, set for each absorption line. The results of \(\alpha\) are graphically plotted in Figure~\ref{fig:alpha_Results_quantumnum}. 
The median of the inferred \(\alpha\) is 0.99, with most transitions --excluding \(J_\mathrm{lower} = 20\)--falling within the range of 0.5 to 1.5. This suggests that the line strengths provided by HITEMP are accurate within a range of several tens of percent. This fact also justifies the use of HITEMP line strengths for calculating the weak lines.
For the lines with \(J_\mathrm{lower} = 20\), the signal-to-noise ratio is lower at lower temperatures due to weaker absorption, which likely led to greater variability in the inference.

%γrefの比較(X軸：Jlower)
\begin{figure*}
\centering
\includegraphics[width=0.49\linewidth]{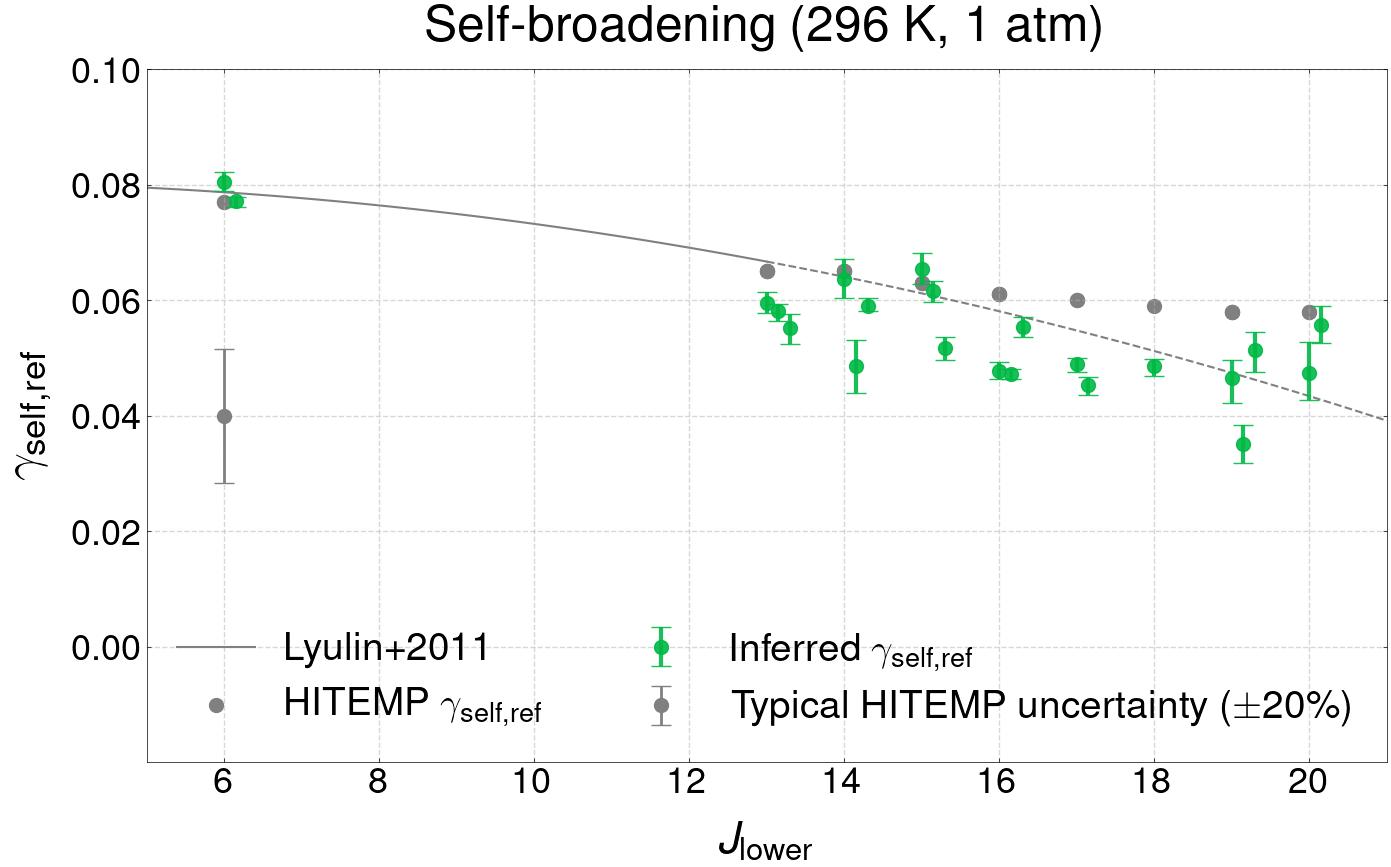}
\includegraphics[width=0.49\linewidth]{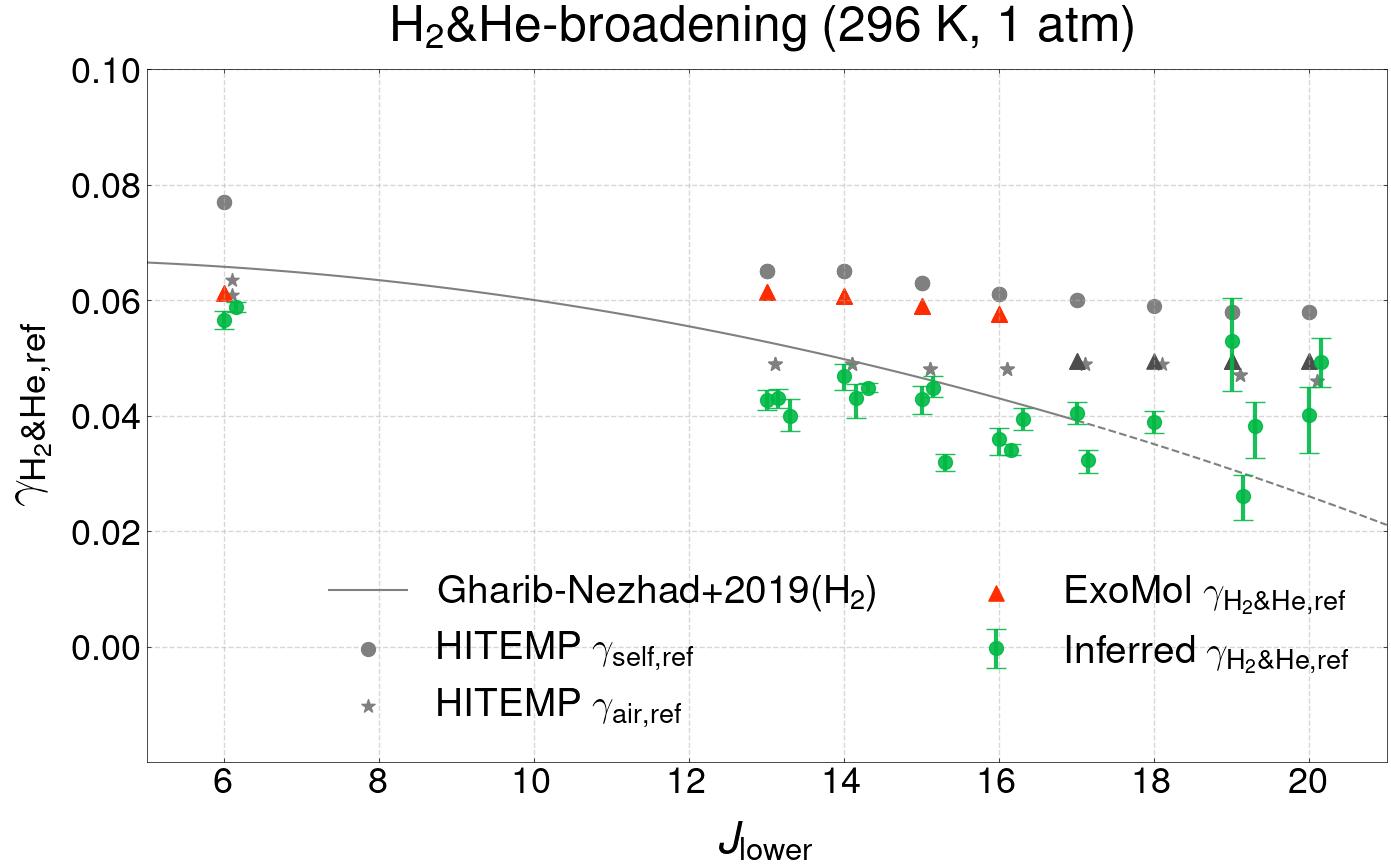}
\caption{Left: Inference results for \(\gamma_{\text{self},\text{ref}}\) across quantum numbers. The green markers with error bars represent the median values and the 1$\sigma$ credible intervals. The gray circles represent \(\gamma_{\text{self},\text{ref}}\) values from HITEMP, and the gray line indicates the empirical equation from \cite{lyulin_measurements_2011}, with extrapolation beyond \(J_\text{lower} \geq \)14 shown as a dashed line assuming the R-branch. 
Right: Results for \(\gamma_{\text{H}_2\&\text{He}, \text{ref}}\). Circles and stars indicate \(\gamma_{\text{self},\text{ref}}\) and \(\gamma_{\text{air},\text{ref}}\) values from HITEMP, while triangles represent \(\gamma_{\text{H}_2\&\text{He},\text{ref}}\) values derived from ExoMol \(\alpha_{\text{H}_2, \text{ref}}\) and \(\alpha_{\text{He}, \text{ref}}\). The gray line corresponds to the empirical equation from \cite{gharib-nezhad_h2-induced_2019} for \(\text{H}_2\) broadening in the \(\lambda\) = 3.3~{\textmu}m absorption band, with extrapolation as a dashed line for \(J_\text{lower} \geq 17\). Since \(\alpha_{\text{H}_2, \text{ref}}\) and \(\alpha_{\text{He}, \text{ref}}\) data extend only up to \(J_\text{lower} = 16\), plots beyond this use default values. For clarity, an offset is applied when multiple lines correspond to the same \(J_{\text{lower}}\).}
\label{fig:Gamma-quantumplot}
\end{figure*}

%nの比較(X軸：Jlower)
\begin{figure*}
\centering
\includegraphics[width=0.49\linewidth]{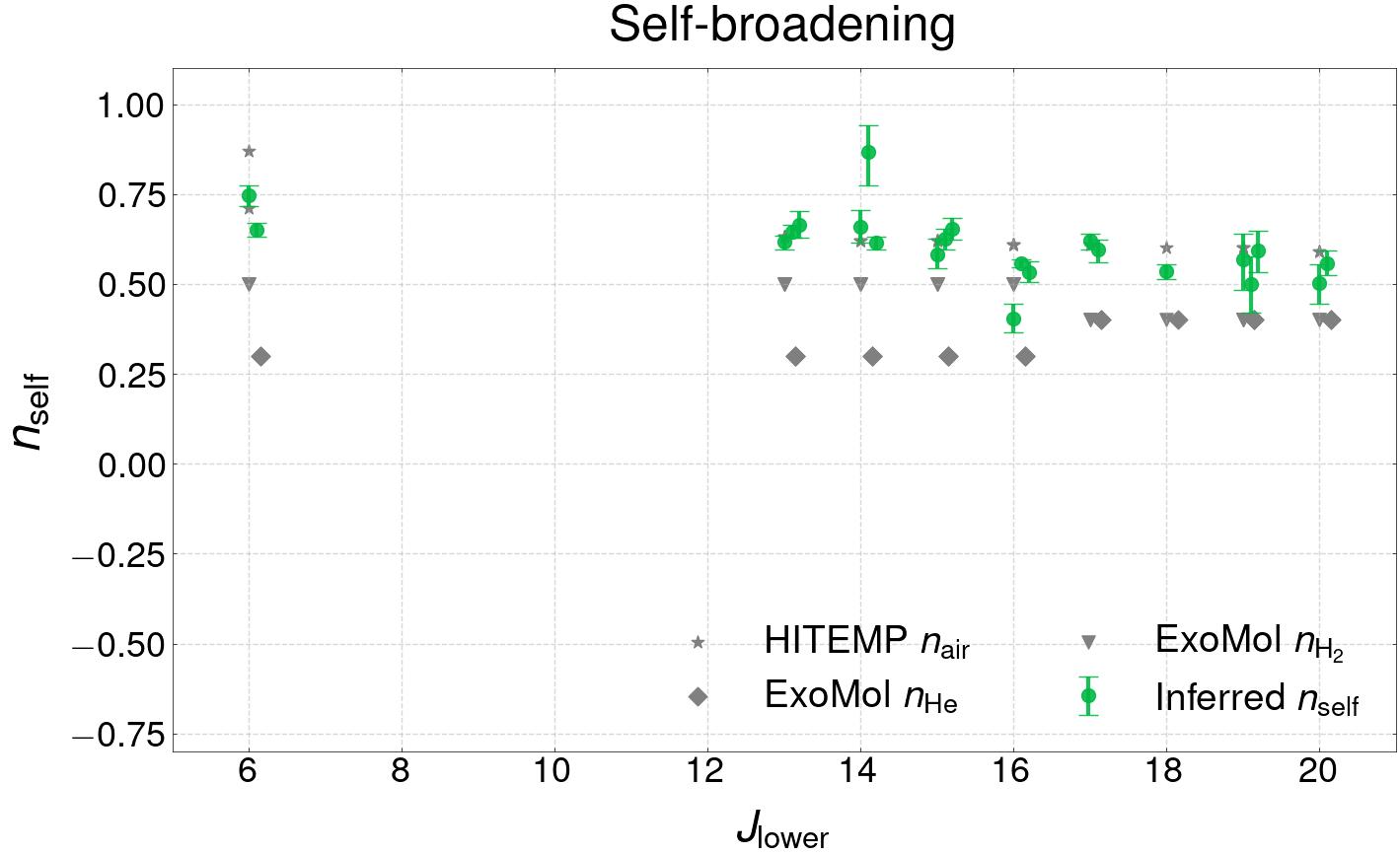}
\includegraphics[width=0.49\linewidth]{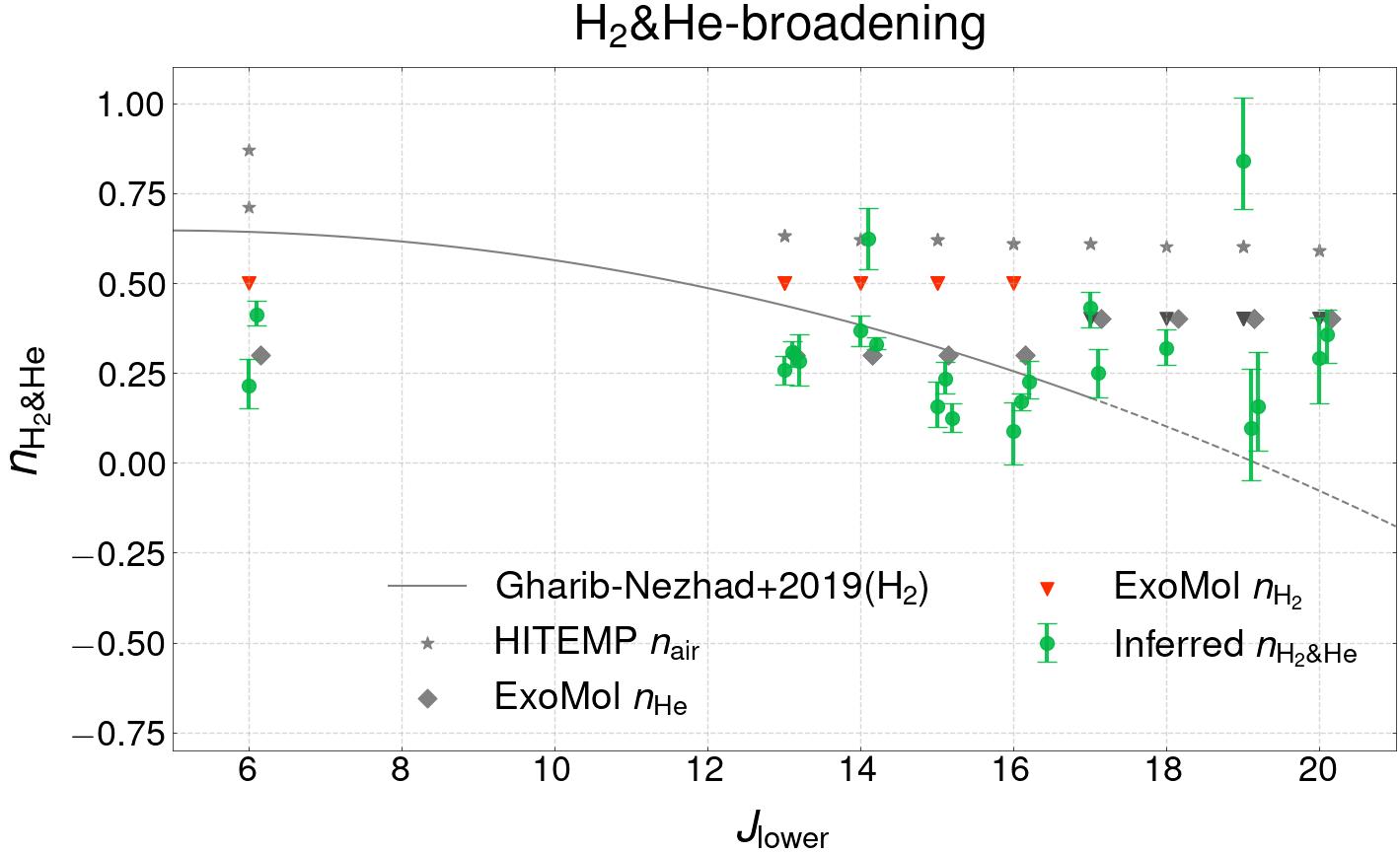}
\caption{Inference results for temperature exponents \( n_{\text{self}} \) (Left) and \( n_{\text{H}_2\&\text{He}} \) (Right) across quantum numbers. The symbols are as follows: stars denote \( n_{\text{air}} \) values from HITEMP, triangles indicate \( n_{\text{H}_2} \), and diamonds represent \( n_{\text{He}} \) from ExoMol. The gray line in the \( n_{\text{H}_2\&\text{He}} \) plot corresponds to the empirical equation from \cite{gharib-nezhad_h2-induced_2019} for \(\text{H}_2\) broadening in the \(\lambda\) = 3.3~{\textmu}m absorption band, with extrapolation shown as a dashed line for \( J_\text{lower} > 17 \).
Note that for \( J_\text{lower} \geq 17 \), the ExoMol values for \( n_{\text{H}_2} \) and \( n_{\text{He}} \) use a “default” value due to the absence of corresponding \(\text{H}_2\) and He broadening data.}
\label{fig:n_Texp-quantumplot}
\end{figure*}

\begin{deluxetable*}{lllllllll}
\tabletypesize{\small}
\tablecaption{Inferred broadening parameters and the scale factors of line strength \(\alpha\) with errors (1$\sigma$ credible intervals of the posterior sample). ID is the identifier of each inferred range from a shorter wavelength. Note that the line center and $J_\text{lower}$ are from HITEMP.}
\label{tab:inference_result_wav}
\tablehead{ID& Wavelength & Wavenumber & $J_\text{lower}$ &$\gamma_{\text{self}, \text{ref}}$ & 
$\gamma_{\text{H}_2\&\text{He}, \text{ref}}$ & $n_{\text{self}, \text{ref}}$ & 
$n_{\text{H}_2\&\text{He}, \text{ref}}$ &$\alpha$\\
& nm &$\text{cm}^{-1}$ && $\text{cm}^{-1}$/atm & $\text{cm}^{-1}$/atm &&&}
\startdata
1 & 1613.68663 & 6196.99006 & 20 & $0.056^{+0.003}_{-0.003}$ & $0.049^{+0.004}_{-0.004}$ & $0.56^{+0.04}_{-0.03}$ & $0.36^{+0.07}_{-0.08}$ & $1.74^{+0.10}_{-0.11}$ \\
1 & 1613.71439 & 6196.88347 & 20 & $0.047^{+0.005}_{-0.005}$ & $0.040^{+0.005}_{-0.007}$ & $0.50^{+0.05}_{-0.06}$ & $0.29^{+0.11}_{-0.13}$ & $3.02^{+0.29}_{-0.43}$ \\
2 & 1615.07413 & 6191.66625 & 19 & $0.046^{+0.003}_{-0.004}$ & $0.053^{+0.007}_{-0.009}$ & $0.57^{+0.07}_{-0.09}$ & $0.84^{+0.18}_{-0.14}$ & $0.72^{+0.03}_{-0.04}$ \\
3 & 1615.75394 & 6189.06120 & 19 & $0.051^{+0.003}_{-0.004}$ & $0.038^{+0.004}_{-0.006}$ & $0.59^{+0.06}_{-0.06}$ & $0.16^{+0.15}_{-0.12}$ & $0.85^{+0.02}_{-0.02}$ \\
3 & 1615.78587 & 6188.93889 & 19 & $0.035^{+0.003}_{-0.003}$ & $0.026^{+0.004}_{-0.004}$ & $0.50^{+0.08}_{-0.08}$ & $0.10^{+0.16}_{-0.15}$ & $0.66^{+0.03}_{-0.02}$ \\
4 & 1617.62785 & 6181.89157 & 18 & $0.049^{+0.001}_{-0.002}$ & $0.039^{+0.002}_{-0.002}$ & $0.53^{+0.02}_{-0.02}$ & $0.32^{+0.05}_{-0.05}$ & $1.16^{+0.03}_{-0.03}$ \\
5 & 1619.70477 & 6173.96465 & 17 & $0.045^{+0.001}_{-0.002}$ & $0.032^{+0.002}_{-0.002}$ & $0.59^{+0.03}_{-0.03}$ & $0.25^{+0.07}_{-0.07}$ & $0.99^{+0.02}_{-0.02}$ \\
5 & 1619.72283 & 6173.89580 & 17 & $0.049^{+0.001}_{-0.001}$ & $0.041^{+0.002}_{-0.002}$ & $0.62^{+0.02}_{-0.03}$ & $0.43^{+0.04}_{-0.05}$ & $1.08^{+0.02}_{-0.02}$ \\
6 & 1621.84132 & 6165.83131 & 16 & $0.055^{+0.002}_{-0.002}$ & $0.039^{+0.002}_{-0.002}$ & $0.53^{+0.03}_{-0.03}$ & $0.22^{+0.06}_{-0.04}$ & $1.37^{+0.05}_{-0.05}$ \\
6 & 1621.85218 & 6165.79002 & 16 & $0.047^{+0.001}_{-0.001}$ & $0.034^{+0.001}_{-0.001}$ & $0.56^{+0.01}_{-0.01}$ & $0.17^{+0.02}_{-0.02}$ & $0.92^{+0.02}_{-0.02}$ \\
6 & 1621.85931 & 6165.76292 & 16  & $0.048^{+0.002}_{-0.001}$ & $0.036^{+0.002}_{-0.003}$ & $0.40^{+0.04}_{-0.04}$ & $0.09^{+0.08}_{-0.09}$ & $1.04^{+0.05}_{-0.06}$ \\
7 & 1623.23403 & 6160.54112 & 15 & $0.062^{+0.002}_{-0.002}$ & $0.045^{+0.002}_{-0.002}$ & $0.63^{+0.03}_{-0.03}$ & $0.23^{+0.05}_{-0.04}$ & $0.99^{+0.02}_{-0.02}$ \\
7 & 1623.30589 & 6160.26842 & 15 & $0.065^{+0.003}_{-0.003}$ & $0.043^{+0.002}_{-0.003}$ & $0.58^{+0.05}_{-0.04}$ & $0.16^{+0.07}_{-0.06}$ & $1.09^{+0.03}_{-0.03}$ \\
8 & 1624.05350 & 6157.43263 & 15 & $0.052^{+0.002}_{-0.002}$ & $0.032^{+0.001}_{-0.002}$ & $0.65^{+0.03}_{-0.03}$ & $0.12^{+0.04}_{-0.04}$ & $0.70^{+0.02}_{-0.02}$ \\
9 & 1625.62522 & 6151.47936 & 14 & $0.049^{+0.004}_{-0.005}$ & $0.043^{+0.002}_{-0.003}$ & $0.87^{+0.08}_{-0.09}$ & $0.62^{+0.08}_{-0.08}$ & $0.97^{+0.05}_{-0.05}$ \\
9 & 1625.63332 & 6151.44872 & 14 & $0.064^{+0.003}_{-0.003}$ & $0.047^{+0.002}_{-0.002}$ & $0.66^{+0.05}_{-0.04}$ & $0.37^{+0.04}_{-0.04}$ & $1.33^{+0.05}_{-0.06}$ \\
10 & 1626.13977 & 6149.53289 & 14 & $0.059^{+0.001}_{-0.001}$ & $0.045^{+0.001}_{-0.001}$ & $0.61^{+0.02}_{-0.02}$ & $0.33^{+0.02}_{-0.01}$ & $1.06^{+0.02}_{-0.02}$ \\
11 & 1627.17009 & 6145.63902 & 6 & $0.077^{+0.001}_{-0.001}$ & $0.059^{+0.001}_{-0.001}$ & $0.65^{+0.02}_{-0.02}$ & $0.41^{+0.04}_{-0.03}$ & $0.98^{+0.01}_{-0.01}$ \\
11 & 1627.21411 & 6145.47277 & 6 & $0.080^{+0.002}_{-0.002}$ & $0.056^{+0.002}_{-0.001}$ & $0.75^{+0.03}_{-0.03}$ & $0.21^{+0.08}_{-0.06}$ & $0.97^{+0.02}_{-0.03}$ \\
12 & 1628.20455 & 6141.73445 & 13 & $0.055^{+0.002}_{-0.003}$ & $0.040^{+0.003}_{-0.003}$ & $0.66^{+0.04}_{-0.04}$ & $0.28^{+0.07}_{-0.07}$ & $0.75^{+0.03}_{-0.03}$ \\
12 & 1628.21527 & 6141.69404 & 13 & $0.058^{+0.001}_{-0.002}$ & $0.043^{+0.002}_{-0.002}$ & $0.65^{+0.02}_{-0.02}$ & $0.31^{+0.03}_{-0.04}$ & $1.11^{+0.03}_{-0.03}$ \\
12 & 1628.28618 & 6141.42658 & 13 & $0.059^{+0.002}_{-0.002}$ & $0.043^{+0.002}_{-0.002}$ & $0.62^{+0.02}_{-0.02}$ & $0.26^{+0.04}_{-0.04}$ & $0.99^{+0.03}_{-0.03}$ \\
\enddata
\end{deluxetable*}

The left panels in Figure~\ref{fig:Gamma-quantumplot} and \ref{fig:n_Texp-quantumplot} show the comparison of the inferred values of the self-broadening parameter \(\gamma_{\text{self},\text{ref}} \) and \(n_{\text{self}}\) at the reference temperature and pressure with the values of two databases.
Most of the median value for the inferred lines (16/22 lines) shows a deviation of more than 1$\sigma$ from HITEMP \(\gamma_{\text{self},\text{ref}} \), which is 10--40\% lower. In addition, the values are inferred within the error of 2--12\%. Our measurements are more precise than the values provided in HITEMP because the uncertainties for these lines are classified as ``$\ge$ 20\%'' HITEMP \(\gamma_{\text{self}}\). Our measured temperature exponents agree well with those ($n_\mathrm{air}$) in HITEMP although $n_\mathrm{air}$ is not for the self-broadening but for Earth air.

Our measurements of the $\text{H}_2$/$\text{He}$ broadening parameters \(\gamma_{\text{H}_2\&\text{He}, \text{ref}}\) and their exponent $n_{\text{H}_2\&\text{He}}$ are shown in the right panels of Figures~\ref{fig:Gamma-quantumplot} and \ref{fig:n_Texp-quantumplot}. 
The inferred \(\gamma_{\text{H}_2\&\text{He}, \text{ref}}\) values are within the error of \(\pm\)2--20\% in a 1$\sigma$ credible interval.
All of our measured hydrogen/helium broadening parameters show values that are 5--45\% lower than those in ExoMol ($\text{H}_2$/He). In ExoMol, the broadening parameters data for \(\text{H}_2\) and He is restricted to \(J_{\text{lower}} \le 16\). For the lines with quantum numbers exceeding this range, we use the ``default'' constant value instead. 
The measured temperature exponents do not depend much on $J_\mathrm{lower}$, and are almost constant around 0.27. These values are milder than those for $\text{H}_2$ and close to the value of He provided in ExoMol ($n_{\text{H}_2} \sim$ 0.5, $n_{\text{He}} \sim$ 0.3).

\subsection{Dependence of Pressure Broadening on the Quantum Number J}

It is recognized that the broadening parameter \(\gamma_\text{ref}\) exhibits a general correlation with the quantum number of the lower state \( J_{\text{lower}} \), with a slight decrease as the quantum number increases regardless of the broadening molecule (e.g. for self-broadening: \cite{lyulin_measurements_2011}, for H\(_2\)-broadening:  \cite{gharib-nezhad_h2-induced_2019}).
In Figure~\ref{fig:Gamma-quantumplot} and \ref{fig:n_Texp-quantumplot}, we illustrate the broadening parameters as a function of \( J_{\text{lower}} \), at the reference pressure (1 atm) and temperature (296~K). 
\cite{lyulin_measurements_2011} suggested the empirical function for \(\gamma_{\text{self}}\) associated with the measured line within the range \(0 \le |m| \le 14\) (For P- and Q-branches \(m=J_\text{lower}\), and for the R-branch \(m=J_\text{lower}+1\). inferred lines are all R-branch) is expressed as follows:
\begin{equation}
\begin{split}
\gamma_{\text{self},\text{ref}} &= 0.080327 \\
&+ 2.432 \times 10^{-4} |m| - 1.1519 \times 10^{-4} |m|^2
\label{eq:gself_quantum_Lyulin2011}
\end{split}
\end{equation}
and there is no clear correlation between \(n_{\text{self}}\) and \(m\), they derived the mean \(n_{\text{self}}\) = 0.84.
In the left panel in Figure~\ref{fig:Gamma-quantumplot}, the solid line represents the empirical relation (Equation~\ref{eq:gself_quantum_Lyulin2011}), while the dashed segment of the line is its extrapolation of the relation to higher quantum numbers. Our measurements of the self-broadening parameters show values that are generally about 10\% lower than the extrapolated empirical relation, yet the dependency follows a similar trend.

\cite{gharib-nezhad_h2-induced_2019} suggested the empirical relations for \(\gamma_{\text{H}_2, \text{ref}}\) and \(n_{\text{H}_2}\) as functions of \(J_{\text{lower}}\):
\begin{equation}
\begin{split}
\gamma_{\text{H}_2, \text{ref}} &= 0.066 \\
&+ 0.0008 \times   J_{\text{lower}} - 0.00014 \times J_{\text{lower}}^2,
\label{eq:gH2_quantum_gharib2019}
\end{split}
\end{equation}
and
\begin{equation}
\begin{split}
n_{\text{H}_2} &= 0.570 \\
&+ 0.0310 \times  J_{\text{lower}} - 0.00317 \times  J_{\text{lower}}^2.
\label{eq:nH2_quantum_gharib2019}
\end{split}
\end{equation}
It should be noted that our background atmosphere is a mixture of hydrogen and helium, whereas their relations were derived using a pure hydrogen background.
As shown in the right panel of Figure~\ref{fig:Gamma-quantumplot}, the trend observed in \(\gamma_{\text{H}_2\&\text{He}, \text{ref}}\) across quantum numbers aligns with the empirical relation and its extrapolation. 
In contrast, the measured temperature exponents for the hydrogen/helium atmosphere remain almost constant rather than decreasing as a function of $J_{\text{lower}}$. 

\section{Discussion and Summary}\label{sec:summary}
\subsection{Impact on exoplanets and substellar atmospheres}
\begin{figure}
\centering
\includegraphics[width=\linewidth]{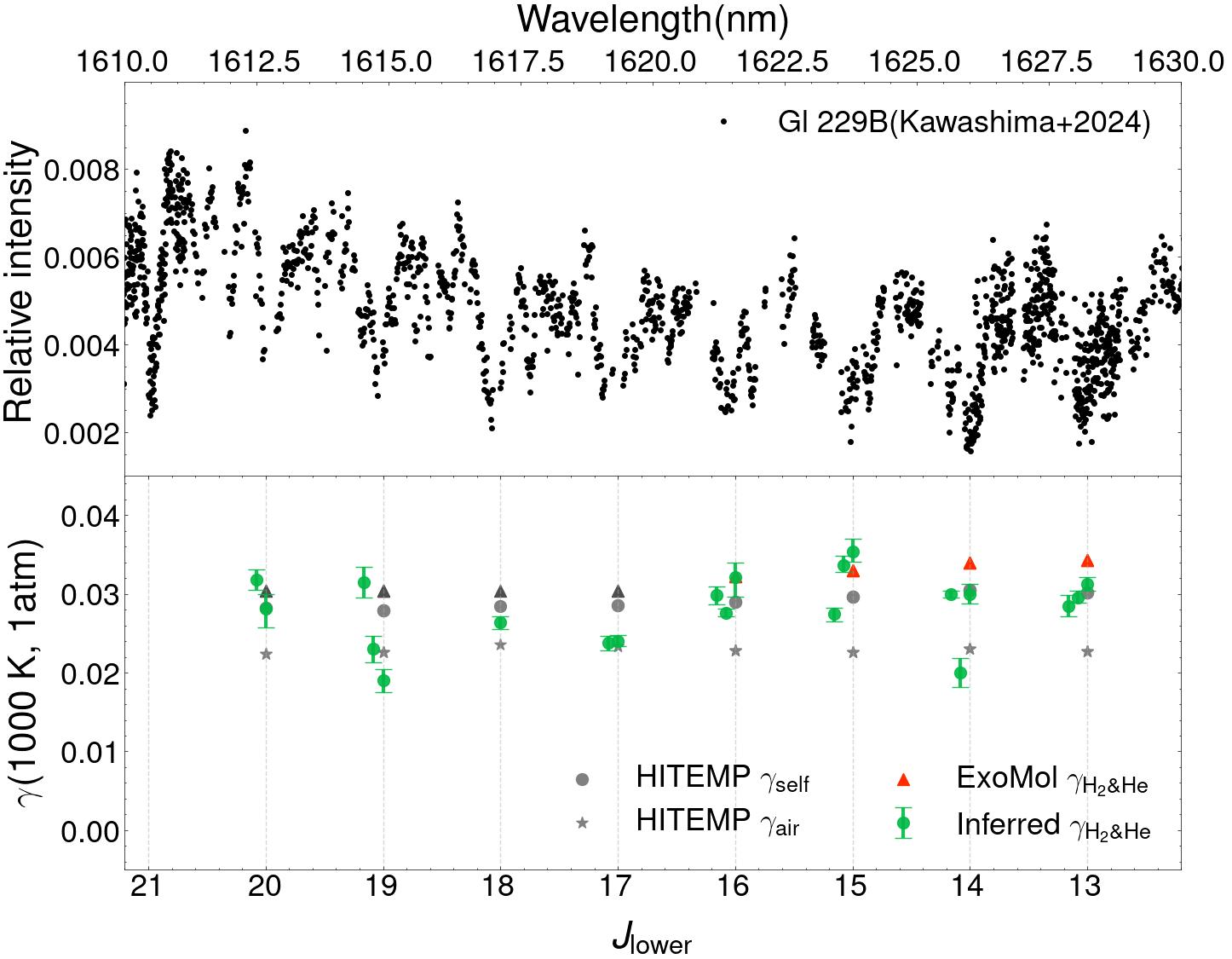}
\caption{Top: high-resolution spectrum of the T dwarf Gl229B as observed by \cite{2024arXiv241011561K}.
Bottom: Inference results of \(\gamma_{\text{H}_2\&\text{He}}(P,T)\) at $P=$ 1 atm, $T=$ 1000~K with quantum numbers. The symbols are as follows: circles represent \(\gamma_{\text{self}}(P,T)\) and stars denote \(\gamma_{\text{air}}(P,T)\) derived from HITEMP, triangles indicate \(\gamma_{\text{H}_2\&\text{He}}(P,T)\) from ExoMol.  Note that the ExoMol \(\gamma_{\text{H}_2\&\text{He}}(P,T)\) for \(J_\text{lower} \ge 17\) is using the default value.}
\label{fig:Gamma_Results_obsGL229B}
\end{figure}

To apply our experimental results to the atmospheres of brown dwarfs and hot gaseous exoplanets, we examine the findings from the previous section to understand broadening effects at higher temperatures, rather than at 296~K. 
Figure~\ref{fig:Gamma_Results_obsGL229B} shows the actual brown dwarf spectrum of the T-dwarf, Gl229B \citep{2024arXiv241011561K}, alongside the measured $\gamma_{\text{H}_2\&\text{He}}(P,T)$ at $P$ = 1 atm and $T$ = 1000~K for the corresponding R-branches to the strong lines in the top panel. These values were converted using the HMC samplings of $\gamma_{\text{H}_2\&\text{He}, \text{ref}}$ (i.e. broadening parameters at the reference temperature 296K) and $n_{\text{H}_2\&\text{He}}$ as described in the previous section. 

Although the broadening values under a hydrogen/helium atmosphere at the reference temperature are systematically lower than those provided by ExoMol (\S 4.2), the inferred widths at 1000~K are close to the ExoMol values for \( J_\mathrm{lower} \le 16 \), due to the mild temperature exponents. For \( J_\mathrm{lower} \geq 17 \), ExoMol does not provide broadening parameters for a hydrogen/helium background, and its “default” value (0.029) were almost along with the inferred \( \gamma_{\text{H}_2\&\text{He}}(P,T)\) for \(J_\mathrm{lower}\) = 17--20 by coincidence, although generally 20\% higher than the inferred values at $T$ = 296K. However, it means that the deviation becomes larger at lower temperatures. While we have not measured all of the strong lines in the H-band, we suggest using \( n_{\text{H}_2\&\text{He}} = 0.27 \) and \( \gamma_{\text{H}_2\&\text{He}, \text{ref}} = 0.040 \) to approximate our measured values in practical analyses of substellar objects.

%γH2He(1000K,1atm)とGL229Bの比較

\subsection{Future prospects}
In this paper, we experimentally investigated methane pressure broadening in a hydrogen-helium atmosphere at high temperatures ($\sim$ 1000~K). However, other key molecules for the atmospheres of brown dwarfs and hot gaseous planets include CO and H\(_2\)O. CO, in particular, due to its diatomic structure, has a simple line structure with fewer line mixtures in the R and P branches, making pressure broadening especially significant. For example, \cite{kawahara_autodifferentiable_2022} used high-resolution spectra of CO in the 2.3 micron band to retrieve the mass of one of the brown dwarf binary Luhman 16A by comparing spectral retrieval results with dynamical mass measurements from astrometry. They found that differences in the pressure broadening parameters for CO (from HITEMP and ExoMol) led to variations in the mass inference. Accurately using CO pressure broadening parameters for mass inference from high-dispersion spectra alone could enable mass inference in systems where dynamical mass is unavailable, such as single brown dwarfs or low-mass gaseous planets and substellar companions. Consequently, we plan to use the measurement setup established in this study to investigate CO pressure broadening in hydrogen-helium atmospheres at high temperatures.

\subsection{Summary}
In this paper, we experimentally measured the pressure broadening of methane in a high temperature (up to 1000~K) hydrogen-helium background atmosphere using glass cells filled with methane in a hydrogen-helium atmosphere. We obtained high-resolution spectra at four temperatures (297, 500, 700, and 1000~K) in the wavelength range of 1.60--1.63~{\textmu}m using a tunable laser. 
A full Bayesian analysis was conducted using the differentiable spectral model {\sf ExoJAX} and HMC-NUTS to infer the pressure broadening for 20 strong methane lines in the R-branch of the $2\nu_3$ band ($J_\mathrm{lower} = $13--20) and 2 lines of $J_\mathrm{lower} = 6$ for an unknown vibration mode. The results indicated that the measured temperature exponent and the reference width of $\mathrm{H_2}$/He pressure broadening are approximately 0.27 and 0.040, which are lower than the values provided by the molecular database ExoMol.
This study emphasizes the need for further measurements of hydrogen-helium broadening at high temperatures across more lines and also more molecular species, such as CO, to improve atmospheric retrievals for substellar objects.

% ACKNOWLEDGEMENT  
We thank Yui Kasagi, Hiroyuki Tako Ishikawa, Akemi Tamanai, Takahiro Koyama, Nami Sakai, Eiichi Tajika, Kazumi Kashiyama, Masahiro Ikoma, Tetsuya Taguchi, Ken Goto, the safety committe of ISAS/JAXA for the fruitful discussions and the kind advice. This experiment was conducted at the Akiruno Experiment Lab of ISAS/JAXA. We appreciate the cooperation of all those involved.
This study was supported by JSPS KAKENHI grant nos. 23K25920 (T.K.) 21H04998, 23H00133, 23H01224 (H.K.), 21K13984, 21H04998, 22H05150, and 23H01224 (Y.K.), The Mitsubishi Foundation (202310018), and the PROJECT Research from the Astrobiology Center (AB0505), and JST SPRING, Grant Number JPMJSP2104(K.H.), the TOBE MAKI Scholarship Foundation(K.H.). Numerical computations were in part carried out on the GPU system at Center for Computational
Astrophysics, National Astronomical Observatory of Japan.

\software{corner.py \citep{corner}, ExoJAX \citep{kawahara_autodifferentiable_2022, 2024arXiv241006900K}, JAX \citep{jax2018github}, matplotlib \citep{Hunter:2007}, numpy \citep{harris2020array}, NumPyro \citep{phan2019composable}, pandas \citep{reback2020pandas}, SciPy \citep{virtanen_scipy_2020}}

Data availability: Analysed spectral data and the codes for this work can be found at \sf{GitHub}. \url{https://github.com/KoHosokawa/Gascell_Exojax}
%\clearpage

\appendix
\section{Inteferometric fringe model}
\label{Frindge_removal}
\begin{figure}
\centering
\includegraphics[width=0.6\linewidth]{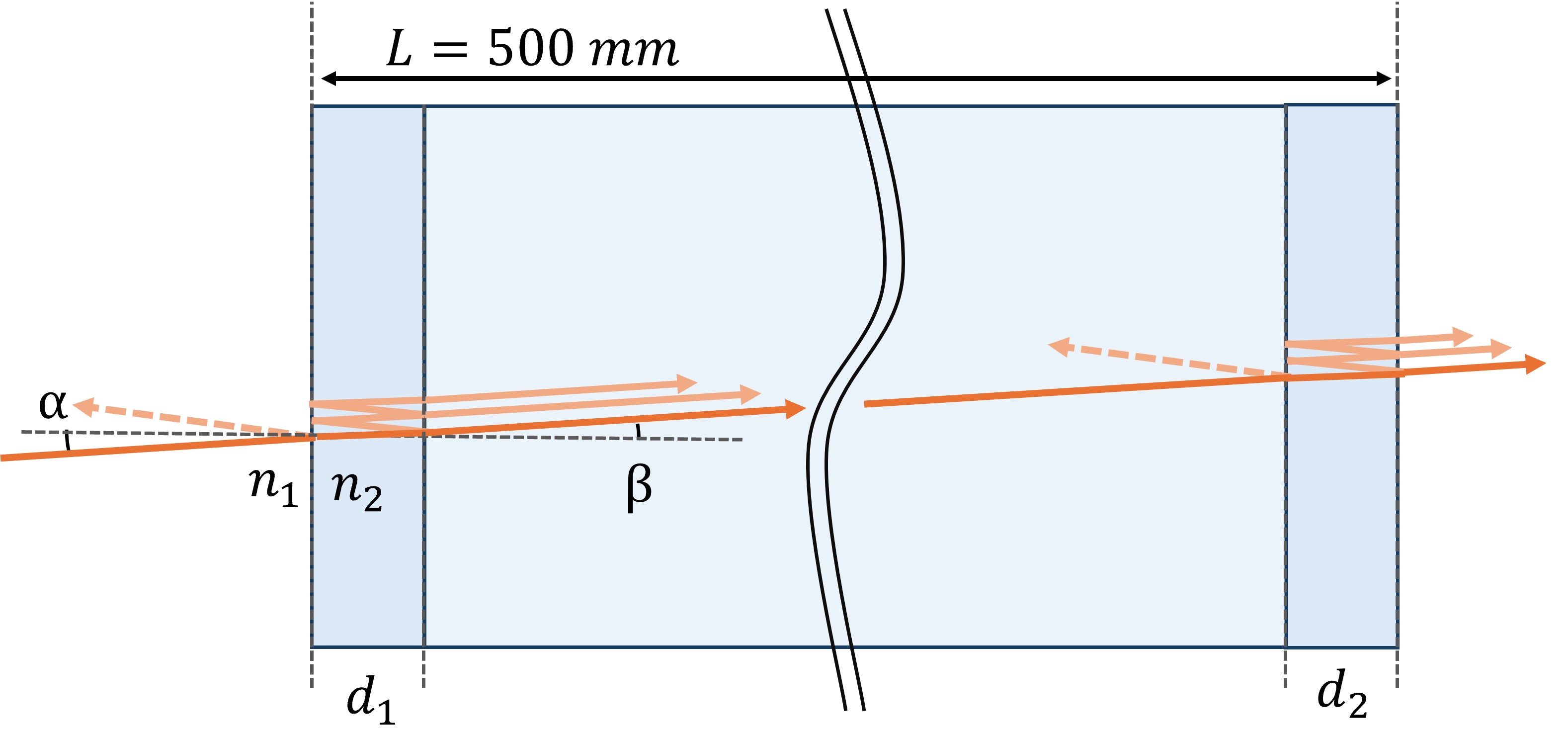}
\caption{Illustration of the transmission through the cell}
\label{fig:Gascell}
\end{figure}

As shown in Fig.~\ref{fig:Gascell}, we consider a plane-parallel glass plate with thickness d and refractive index $n_{2}$ placed in the air with refractive index $n_{1}$. Light enters at an angle of incidence $\alpha$ and refracts at an angle $\beta$. The light undergoes multiple reflections between the front and back surfaces of the glass, causing interference that leads to wavelength-dependent variations in the intensity of transmitted light, resulting in interference fringes. The transmittance $T$ of the light under these conditions is given by the following expression \citep{Hecht}

\begin{equation}
T = \frac{1}{1+ \frac{4R}{(1-R)^2} \sin ^2 (\delta /2)}
\end{equation}

In this equation, $\delta$ represents the phase difference between two adjacent light beams passing through the glass and can be written as follows:

\begin{equation}
\delta = \frac{4\pi n_{2} d \cos \beta}{\lambda}
\end{equation}

Here $\lambda$ is the wavelength, $R$ is the Fresnel reflection coefficient for a single reflection on the glass surface, which can be expressed as follows for $S$ and $P$ polarization respectively:

\begin{equation}
R_{s} = \bigg(\frac{n_{1} \cos \alpha-n_{2} \cos \beta}{n_{1} \cos \alpha+n_{2} \cos\beta} \bigg)^{2}, 
R_{p} = \bigg(\frac{n_{1} \cos \beta-n_{2} \cos \alpha}{n_{1} \cos \beta+n_{2} \cos\alpha} \bigg)^{2}
\end{equation}

The refractive index $n$ for fused silica can be expressed as follows from \cite{malitson_interspecimen_1965}. 

\begin{equation}
\label{eq:fused_silica_index}
n = \sqrt{1+ \frac{0.6961663 \lambda^2}{\lambda^{2}-0.0684043^{2}}+\frac{0.4079426\lambda^{2}}{\lambda^{2}-0.1162414^{2}}+\frac{0.8974794 \lambda^{2}}{\lambda^{2}-9.896161^{2}}}
\end{equation}

Note that the unit of $\lambda$ in Equation~\ref{eq:fused_silica_index} is {\textmu}m. In the case of gas cell windows consisting of two parallel glass plates with thicknesses \(d_1\) and \(d_2\) that are positioned apart, the total transmittance \(T_{all}\) after the light passes through both glass plates is expressed as follows:

\begin{equation}
T_{all} = T_{1} \times T_{2}
\end{equation}

where $T_1$ and $T_2$ are the transmittance through each glass plate. To simplify the problem, we assumed that the light transmitted through the first glass plate and the light reflected by the second glass plate do not interfere with each other.

\section{Comparison with the conventional fitting method}
In order to examine the difference in results due to the varying estimation methods, we also performed the estimation of the broadening parameters with the nonlinear least-squares method, which is mainly used in previous studies~\citep[e.g.,][]{gharib-nezhad_h2-induced_2019, sung_h2-pressure_2020, yousefi_line_2021}. We performed the estimation for the same wavenumber regions and target lines using 8 setup spectra as inferred by Bayesian analysis in Section~\ref{sec:modeling}. The curve\_fit method in {\sf SCiPy}~\citep{virtanen_scipy_2020} was used with the Trust Region Reflective optimization~\citep{branch_subspace_1999} to minimize the following cost function $\mathcal{F}$, performed separately for each wavenumber region $\Lambda$: 
\begin{align}
\mathcal{F}^\supl = \frac{1}{2}\sum_{k=1}^{8} {{\sum_{i} [y^\supkl_i- \mu^\supkl(\nu_i)]^2}},
\end{align}

In contrast to the likelihood function of Bayesian analysis (Equation~\ref{eq:likelihood}), the number of free parameters for this estimation is reduced by the number of $\sigma_{\rm noise}^\supkl$ set for each setup and wavenumber region of the spectra, ranging from 45 to 55 for each wavenumber region. The parameter error was estimated by calculating the standard deviation of the spectra over several wavelength ranges where absorption could not be observed, considering that as the amount of noise for each setup $\sigma_{\rm noise}^{(k)}$, and using the covariance matrix.

The results are shown in Table~\ref{tab:Leastsquare_result_wav}. Approximately 1/3 of the estimated lines (7/22 lines) result in all $\gamma_\text{ref}$, $n$, and $\alpha$ falling within the error (which equals the 1$\sigma$ credible interval) of the results by Bayesian inference in Table~\ref{tab:inference_result_wav}. The deviations ranged from $+$8 to $-$44\% for $\gamma_\text{ref}$, +0.06 to $-$0.27 for $n$, and +6 to $-$42\% for $\alpha$. Only the parameters of the two lines with $J_\text{lower}$ =20 differed more than the 3$\sigma$ error of inference results. There was no systematic offset for $\gamma_\text{ref}$, $n$ and $\alpha$ between the estimated results using the least-squares method and the median of the inference results.
Since the estimation by the least-squares method did not take into account the spectral noise for fitting to the spectra, unlike Bayesian analysis, it is possible that some of the outliers on the spectrum affected the difference of the parameter results. As we did not test every possible initial parameter value, the fits may have stuck in local minima.

Focusing on the $\text{H}_2$/He broadening parameters, the median of \( n_{\text{H}_2\&\text{He}}\) and \( \gamma_{\text{H}_2\&\text{He}, \text{ref}}\) for \( J_\mathrm{lower} \) = 13--20 was 0.26 and 0.039, which is generally consistent with the inference results described in Section~\ref{sec:result}.

\begin{deluxetable*}{lllllllll}
\tabletypesize{\small}

\tablecaption{Result of broadening parameters and the scale factors of line strength \(\alpha\) with 1$\sigma$ errors by least-squares fits. Discription of ID, line center and $J_\text{lower}$ are same with Table~\ref{tab:inference_result_wav}.}
\label{tab:Leastsquare_result_wav}
\tablehead{ID& Wavelength & Wavenumber & $J_\text{lower}$ &$\gamma_{\text{self}, \text{ref}}$ & 
$\gamma_{\text{H}_2\&\text{He}, \text{ref}}$ & $n_{\text{self}, \text{ref}}$ & 
$n_{\text{H}_2\&\text{He}, \text{ref}}$ &$\alpha$\\
& nm &$\text{cm}^{-1}$ && $\text{cm}^{-1}$/atm & $\text{cm}^{-1}$/atm &&&}
\startdata
1 & 1613.68663 & 6196.99006  & 20 & 0.046$\pm$0.005 & 0.040$\pm$0.007 & 0.49$\pm$0.08 & 0.27$\pm$0.16 & 1.39$\pm$0.13 \\ 
1 & 1613.71439 & 6196.88347  & 20  & 0.029$\pm$0.007 & 0.023$\pm$0.008 & 0.32$\pm$0.17 & 0.02$\pm$0.31 & 1.77$\pm$0.33 \\ 
2 & 1615.07413 & 6191.66625  & 19  & 0.043$\pm$0.007 & 0.051$\pm$0.013 & 0.52$\pm$0.14 & 0.81$\pm$0.26 & 0.70$\pm$0.06 \\ 
3 & 1615.75394 & 6189.06120  & 19  & 0.053$\pm$0.005 & 0.037$\pm$0.006 & 0.60$\pm$0.10 & 0.09$\pm$0.18 & 0.89$\pm$0.04 \\ 
3 & 1615.78587 & 6188.93889  & 19  & 0.037$\pm$0.005 & 0.026$\pm$0.006 & 0.51$\pm$0.12 & 0.08$\pm$0.23 & 0.68$\pm$0.04 \\ 
4 & 1617.62785 & 6181.89157  & 18  & 0.046$\pm$0.002 & 0.039$\pm$0.003 & 0.50$\pm$0.04 & 0.30$\pm$0.07 & 1.13$\pm$0.03 \\ 
5 & 1619.70477 & 6173.96465  & 17  & 0.044$\pm$0.002 & 0.033$\pm$0.002 & 0.56$\pm$0.05 & 0.25$\pm$0.08 & 0.99$\pm$0.02 \\ 
5 & 1619.72283 & 6173.89580  & 17  & 0.048$\pm$0.002 & 0.040$\pm$0.002 & 0.62$\pm$0.04 & 0.40$\pm$0.06 & 1.08$\pm$0.02 \\ 
6 & 1621.84132 & 6165.83131  & 16  & 0.057$\pm$0.004 & 0.041$\pm$0.003 & 0.59$\pm$0.07 & 0.25$\pm$0.08 & 1.41$\pm$0.10 \\ 
6 & 1621.85218 & 6165.79002  & 16  & 0.046$\pm$0.002 & 0.033$\pm$0.002 & 0.55$\pm$0.03 & 0.17$\pm$0.04 & 0.92$\pm$0.04 \\ 
6 & 1621.85931 & 6165.76292  & 16  & 0.049$\pm$0.005 & 0.033$\pm$0.004 & 0.46$\pm$0.13 & 0.00$\pm$0.14 & 0.95$\pm$0.10 \\ 
7 & 1623.23403 & 6160.54112  & 15  & 0.061$\pm$0.002 & 0.045$\pm$0.002 & 0.61$\pm$0.04 & 0.25$\pm$0.05 & 0.99$\pm$0.03 \\ 
7 & 1623.30589 & 6160.26842  & 15  & 0.065$\pm$0.003 & 0.046$\pm$0.003 & 0.55$\pm$0.06 & 0.22$\pm$0.07 & 1.12$\pm$0.04 \\ 
8 & 1624.05350 & 6157.43263  & 15  & 0.053$\pm$0.002 & 0.035$\pm$0.001 & 0.66$\pm$0.03 & 0.16$\pm$0.04 & 0.74$\pm$0.02 \\ 
9 & 1625.62522 & 6151.47936  & 14  & 0.038$\pm$0.003 & 0.038$\pm$0.003 & 0.65$\pm$0.10 & 0.59$\pm$0.11 & 0.87$\pm$0.05 \\ 
9 & 1625.63332 & 6151.44872  & 14  & 0.059$\pm$0.003 & 0.045$\pm$0.003 & 0.59$\pm$0.05 & 0.36$\pm$0.05 & 1.30$\pm$0.06 \\ 
10 & 1626.13977 & 6149.53289 & 14  & 0.059$\pm$0.001 & 0.044$\pm$0.001 & 0.60$\pm$0.01 & 0.30$\pm$0.01 & 1.05$\pm$0.01 \\ 
11 & 1627.17009 & 6145.63902  & 6  & 0.079$\pm$0.001 & 0.060$\pm$0.001 & 0.65$\pm$0.03 & 0.43$\pm$0.05 & 1.01$\pm$0.01 \\ 
11 & 1627.21411 & 6145.47277  & 6  & 0.084$\pm$0.002 & 0.060$\pm$0.001 & 0.75$\pm$0.05 & 0.25$\pm$0.09 & 1.03$\pm$0.02 \\ 
12 & 1628.20455 & 6141.73445  & 13  & 0.056$\pm$0.002 & 0.039$\pm$0.001 & 0.62$\pm$0.04 & 0.26$\pm$0.04 & 0.73$\pm$0.02 \\ 
12 & 1628.21527 & 6141.69404  & 13  & 0.063$\pm$0.001 & 0.045$\pm$0.001 & 0.66$\pm$0.02 & 0.33$\pm$0.02 & 1.18$\pm$0.02 \\ 
12 & 1628.28618 & 6141.42658  & 13  & 0.062$\pm$0.001 & 0.043$\pm$0.001 & 0.62$\pm$0.02 & 0.27$\pm$0.02 & 1.01$\pm$0.01 \\ 
\enddata
\end{deluxetable*}

\clearpage

\bibliography{references_zotero, references}{}
\bibliographystyle{aasjournal}

\end{document}